\documentclass[aps,prl,11pt,onecolumn,superscriptaddress]{revtex4-2}
\usepackage{graphicx}
\usepackage{amsmath,amsfonts}
\usepackage{color}
\usepackage{ulem}
\usepackage{braket}
\newcommand{\upsub}[1]{\sb{\mathrm{#1}}}
\newcommand{\upsup}[1]{\sp{\mathrm{#1}}}

\begingroup\lccode`~=`_\lowercase{\endgroup\let~\upsub}
\begingroup\lccode`~=`^\lowercase{\endgroup\let~\upsup}

\AtBeginDocument{%
	\catcode`_=12 \catcode`^=12
	\mathcode`_="8000
	\mathcode`^="8000
}

\begin{document}


	\title{Interface second harmonic generation enhancement in hetero-bilayer van der Waals nanoantennas}

    \author{Andrea Tognazzi\textsuperscript{\S}}
    \email{andrea.tognazzi@unipa.it}
    \thanks{\textsuperscript{\S}These authors contributed equally to this work.}
 	\affiliation{University of Palermo, Department of Engineering, Viale delle Scienze, 90128,Palermo,  Italy} \affiliation{National Institute of Optics – National Research Council (INO-CNR), Via Branze 45, 25123, Brescia, Italy} 
    \author{Paolo Franceschini\textsuperscript{\S}}
     \email{paolo.franceschini@unibs.it}
     \thanks{\textsuperscript{\S}These authors contributed equally to this work.}
 	\affiliation{University of Brescia, Department of Information Engineering, Via Branze 38, 25123, Brescia, Italy} 
    \affiliation{National Institute of Optics – National Research Council (INO-CNR), Via Branze 45, 25123, Brescia, Italy}
	\author{Jonas Biechteler}
    \author{Enrico Baù}
	\affiliation{Chair in Hybrid Nanosystems, Nanoinstitute Munich, Faculty of Physics, Ludwig-Maximilians-Universit{\"a}t M{\"u}nchen, 80539 Munich, Germany}
    \author{Alfonso Carmelo Cino}
    \affiliation{University of Palermo, Department of Engineering, Viale delle Scienze, 90128,Palermo,  Italy}
    \author{Andreas Tittl}
    \affiliation{Chair in Hybrid Nanosystems, Nanoinstitute Munich, Faculty of Physics, Ludwig-Maximilians-Universit{\"a}t M{\"u}nchen, 80539 Munich, Germany}
    \author{Costantino De Angelis}
 	\affiliation{University of Brescia, Department of Information Engineering, Via Branze 38, 25123, Brescia, Italy} 
    \affiliation{National Institute of Optics – National Research Council (INO-CNR), Via Branze 45, 25123, Brescia, Italy}
    \author{Luca Sortino}
    \email{luca.sortino@physik.uni-muenchen.de}
    \affiliation{Chair in Hybrid Nanosystems, Nanoinstitute Munich, Faculty of Physics, Ludwig-Maximilians-Universit{\"a}t M{\"u}nchen, 80539 Munich, Germany}
	\date{\today}
	\maketitle

\noindent
\textbf{Layered van der Waals (vdW) materials have emerged as a promising  platform for nanophotonics due to large refractive indexes and giant optical anisotropy. Unlike conventional dielectrics and semiconductors, the absence of covalent bonds between layers allows for novel degrees of freedom in designing optically resonant nanophotonic structures down to the atomic scale, from the precise stacking of vertical heterostructures to controlling the twist angle between crystallographic axes. Specifically, while transition metal dichalcogenides monolayers exhibit giant second order nonlinear responses, their bulk counterparts with 2H stacking have zero second order response.
In this work, we show second harmonic generation (SHG) arising from the interface of WS$_2$/MoS$_2$ hetero-bilayer thin films with an additional SHG enhancement in nanostructured optical antennas mediated by both the excitonic resonances and the anapole condition. When both conditions are met, we observe up to $10^2$ SHG signal enhancement. Our results highlights vdW materials as a platform for designing unique multilayer optical nanostructures and metamaterial, paving the way for advanced applications in nanophotonics and nonlinear optics.}

\pagebreak
\noindent
\textbf{Introduction}

\noindent
High refractive index dielectric materials, such as silicon, gallium phosphide and III-V semicondutors, have emerged due to their exceptional ability to confine light within nanostructured optical resonators \cite{Kuznetsov2016,Cambiasso2017a,Xu2020}. Unlike plasmonic metallic counterparts, which primarily exploit surface plasmon resonances, dielectric materials leverage Mie resonances, with both electric and magnetic components \cite{Kuznetsov2012}. This unique characteristic allows for the exploration of new degrees of freedom in nanophotonic design, where nanoresonators can be precisely engineered to manipulate the interference between electric and magnetic resonances \cite{Liu2017a}. By tailoring the combination of material properties, geometry, and optical excitation, modulation of the directional emission and non-trivial optical states can be achieved. For instance, non-radiative dark states within resonant dielectric nanostructures \cite{Koshelev2019} result from the interference between different radiative channels, leading to strong confinement of electromagnetic energy connected with a suppression of far-field scattered radiation. In this regard, significant attention has been directed towards anapole states \cite{Miroshnichenko2015} and bound states in the continuum \cite{Kang2023}, finding applications for non-linear optics owing to the increased internal electromagnetic energy in the system and versatile control over the radiated pattern \cite{Carletti2018,Koshelev2020}. The fabrication of conventional dielectric nanoresonators typically relies on the growth of polycrystalline thin films, which suffer from lattice mismatch at heterostructure interfaces, thereby limiting optical quality and the potential for creating useful multilayered structures \cite{Meng2023}. Recently, van der Waals (vdW) materials emerged as a new class of crystals for non-linear optics and nanophotonics \cite{Trovatello2024} promising to overcome current dielectric materials' limitations \cite{Khurgin2022,Vyshnevyy2023}. 

Due to their crystal structure, which features strong in-plane covalent bonds and weak vdW forces between planes, vdW crystals can be mechanically exfoliated into thin crystalline layers on arbitrary substrates, down to single-atom thicknesses. Moreover, the absence of covalent bonds between layers enables the deterministic stacking of multiple layers, forming so-called vdW heterostructures \cite{Novoselov2016}. Owing to their remarkable optical and structural properties, atomically thin two-dimensional (2D) semiconductors, such as transition metal dichalcogenides (TMDCs), have been at the forefront of nanophotonics research in recent years, ranging from integrated components, light-matter coupling, and non-linear optics \cite{Autere2018,Mueller2018}. In this regard, second harmonic generation (SHG) has been a long standing technique for characterizing single and few TMDC layers \cite{Malard2013}, as well as for imaging of mechanical deformations \cite{Mennel2018}, further exhibiting unconventional effects, from quantum interference \cite{Lin2018}, to broadband phase matching \cite{Trovatello2020b} and all-optical modulation \cite{Klimmer2021}. 
Beyond their 2D form, vdW materials thin films ($<$100 nm in thickness) have attracted large attention as new building blocks of integrated nanophotonic structures \cite{Lin2022,Munkhbat2022b,Zotev2022b}. They provide exciting properties for nanoscale dielectric resonators, such as large anisotropy \cite{Ermolaev2020}, high refractive indexes \cite{Munkhbat2022,Zotev2022b}, and wide substrate affinity \cite{Zotev2022}, along with an ever growing library of materials. Additionally, vdW crystals possess the unique property of engineering interface second order non-linear processes \cite{Yao2021} by tuning the twist angle between adjacent layers, opening a novel degree of freedom in the design of non-linear \cite{Kim2023b} and chiral \cite{Voronin2024} vdW-based optical metamaterials. Nanophotonic structures with TMDC thin films have been demonstrated, from anapole nanoantennas \cite{Verre2018, Zotev2022, Zograf2024} and optical metasurfaces \cite{Baranov2020,Nauman2021,Weber2022a,Shen2022b}, to linear and non-linear waveguides \cite{Ling2023,Xu2022a} and optical modulators \cite{Lee2024}. 
As research on TMDCs films for nanophotonics is still in its infancy, so far the exploration of optical resonances in TMDC material nanostructures relied on single exfoliated materials \cite{Sortino2024}. A study on optical resonators made with TMDCs thin film heterostructures is still missing, which could bring to bianisotropic structures, for directional non-linear light emission and surface polaritons \cite{Kruk2022,Basov2020}, or interface driven non-linear effects \cite{Shen1989}, such as the modulation by excitonic resonances of the interface SHG in bulk TMDCs, which, to the best of our knowledge, has not yet been observed. 

\begin{figure}
	\centering
	\includegraphics[width=1\linewidth]{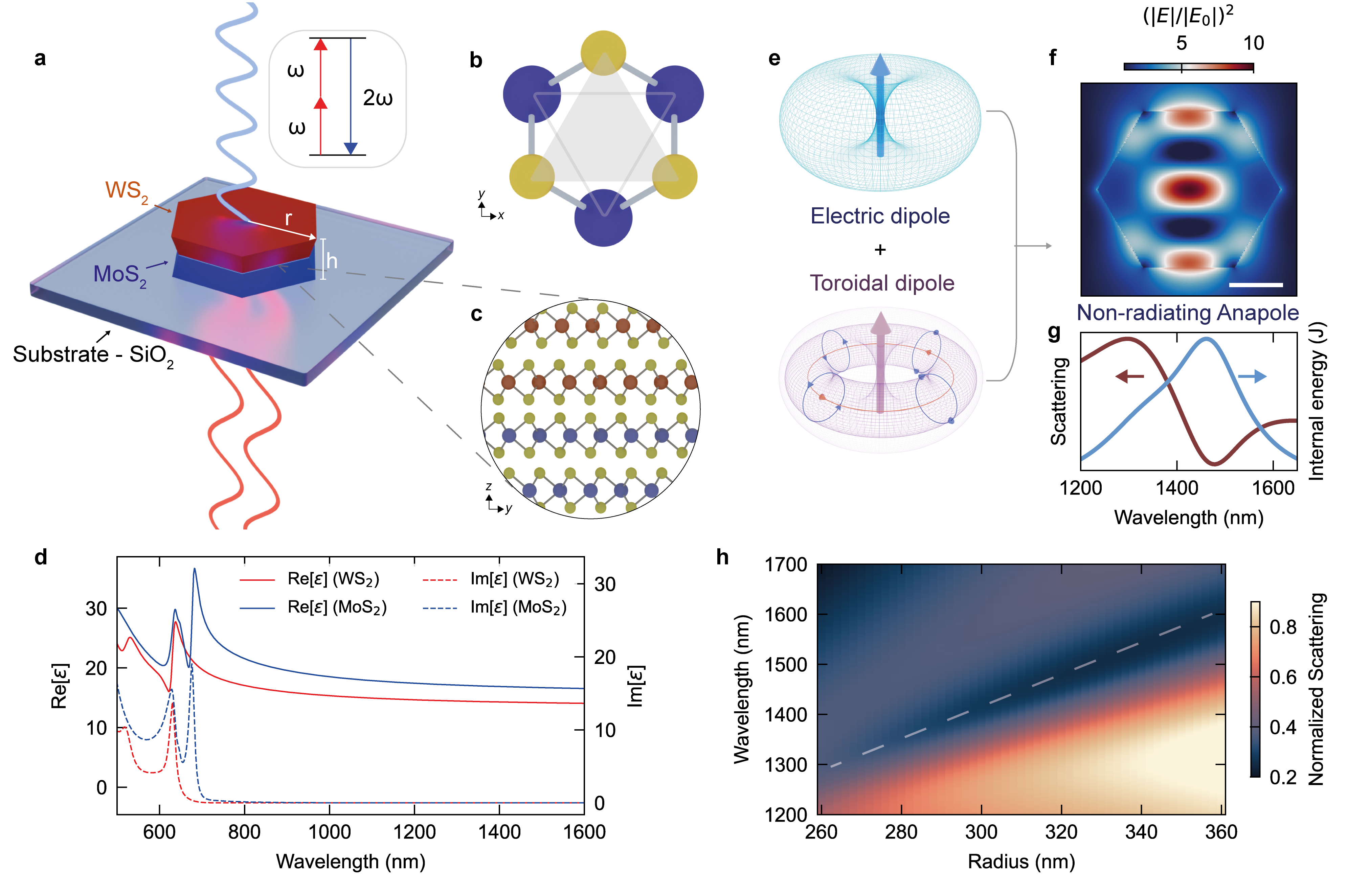}
	\caption{\textbf{Hetero-bilayer WS$_2$/MoS$_2$ van der Waals hexagonal nanoantennas sustaining anapole states.} (a) Schematic of the dual-layer van der Waals (vdW) nanoantenna positioned on a SiO$_2$ substrate, composed of a bottom MoS$_2$ layer (blue) and a top WS$_2$ layer (red), each approximately 100 nm thick. Inset shows the second harmonic generation process where the absorption of two photons at the fundamental frequency $\omega$ leads to the generation of a frequency-doubled photon at $2\omega$. (b) Illustration of the  in-plane TMDC crystalline honeycomb symmetry. Metal atoms are depicted in blue and chalcogenide ones in yellow. (c) Illustration of the interface between the two TMDC layers aligned at zero degrees, showing broken inversion symmetry region resulting in the second order non-linear signal. (d) Real and imaginary parts of the in-plane dielectric function for WS$_2$ (red) and MoS$_2$ (blue). Adapted from Ref.\cite{Munkhbat2022}. (e) Far field illustration of an electric dipole (top) and a toroidal dipole (bottom) whose interference generates the anapole state. (f) Numerical FDTD simulations of the electric field intensity, $(|E|/|E_0|)^{2}$, at the anapole wavelength for a WS$_2$/MoS$_2$ disk with layer thicknesses of 92~nm and 115~nm, respectively, and radius of 280~nm. The data is displayed along the \textit{z}-plane of the TMDCs interface. Scale bar: 200 nm. (g) Numerical FDTD simulations of the normalized scattering cross section (in red) exhibiting a minima, and the normalized internal electromagnetic energy (in blue) exhibiting a maximum at the anapole wavelength. (h) Normalized scattering cross section for a WS$_2$/MoS$_2$ disk on a glass substrate, with radial size from 260~nm to 360~nm, and height of 92~nm for the MoS$_2$ layer and 115~nm for the WS$_2$ one. The sample is illuminated with normal incidence light from the air side. The dashed white line indicates the dip in far-field scattering attributed to the anapole state.}
	\label{fig:fig1}
\end{figure}

In this work, we demonstrate SHG arising from the interface between two dielectric vdW materials, and its additional enhancement due to the interplay of excitonic resonances and the anapole states in dielectric nanoresonators. 
Double-layer optical nanoantennas were prepared from TMDCs WS$_2$/MoS$_2$ hetero-bilayer thin films, obtained through mechanical exfoliation and subsequent deterministic stacking. The resulting heterostructure is then processed with standard nanofabrication methods and shaped into hexagonal resonators, revealing the underlying crystal symmetry of the materials. Due to the close values of the refractive index between the two layer, the resonators act as an homogeneous dielectric medium, allowing the design of single structures sustaining non-radiating anapole states. We confirm the presence of anapole states in the fabricated sample by linear optical reflectance measurements in the near infrared region. 
Although bulk TMDCs do not possess broken inversion symmetry, we observe SHG in the presence of the WS$_2$/MoS$_2$ interface. Moreover, a strongly enhanced SH signal is observed in anapole nanoantennas, which is driven either by the presence of exciton resonances at the second harmonic frequency ($2\omega$), due to increased $\chi^{(2)}$ tensor elements of the material, or by the anapole state, which increases the energy density at the fundamental ($\omega$) frequency. Finally, when the SH of the anapole frequency matches the excitonic resonance, we observe up to two orders of magnitude enhancement of SHG compared to the reference hetero-bilayer.
Our findings highlight the unique potential of vdW materials for designing unprecedented vertically stacked nanophotonic structures with arbitrary materials, opening to precise control over crystal thickness and orientation. Additionally, TMDCs could play a critical role in advancing nonlinear optics and nanophotonics with the interplay of intrinsic excitonic states and photonic resonances, opening new avenues for optically active linear and non-linear materials with tailored optical properties.

\section{Results}
Interfaces play a crucial role in breaking the inversion symmetry, a requirement for a non-zero $\chi^{(2)}$ tensor, overcoming materials' restrictions and promoting the generation of second order non-linear processes even in centrosymmetric crystals. Figure~\ref{fig:fig1}a shows an illustration of the hetero-bilayer TMDC nanoantenna for SHG, where the interface between the two vdW materials promotes the symmetry-breaking condition for the SHG process. The dielectric nanoresonator is made of two bulk TMDCs materials, a bottom layer of MoS$_2$ and a top layer of WS$_2$, with an hexagonal shape owing to the selective anisotropic etching \cite{Zotev2022}. The crystal structure of a single layer of TMDCs is shown in Figure~\ref{fig:fig1}b, where the honeycomb lattice leads to a non-centrosymmetric structure in single layers with $D_{3h}$ point group \cite{Malard2013}. However, bulk TMDCs with 2H stacking lack a broken inversion symmetry as each successive layer is rotated 180 degrees compared to the neighbouring ones, resulting in a $D_{6h}$ point group. As such, the emission of a second order non-linear signal in our sample can be ascribed to the breaking of the inversion symmetry generated at the TMDCs bulk interface (Figure~\ref{fig:fig1}c). In order to maximize the efficiency of the interface SHG signal, we aligned the crystal axes to near zero-degree \cite{Zotev2022b}. 
In Figure~\ref{fig:fig1}d are shown the real and imaginary part of bulk MoS$_2$ and WS$_2$ used in this study. Both materials exhibit high refractive indexes in the visible and near-infrared regions. The resonance peaks observed in the visible range correspond to the excitonic states inherent to each material. 

In resonant dielectric nanostructures, the interference of an electric dipole and a toroidal dipole, depicted in Figure~\ref{fig:fig1}e, leads to the creation of the non-radiative anapole state, confining light and boosting non-linear optical processes \cite{Xu2018a}. Figure~\ref{fig:fig1}f shows the finite-difference time-domain (FDTD) numerical simulation of the electric field intensity $(|E|/|E_{0}|)^2$ of a WS$_2$/MoS$_2$ nanohexagon, where $ E $ is electric field amplitude of the scattered field by the antenna and $ E_0 $ the normally incident field. The field exhibits the characteristic anapole field profile \cite{Miroshnichenko2015}, reaching values of one order of magnitude in field enhancement. As shown in Figure~\ref{fig:fig1}g, the anapole condition results in the suppression of the far field radiation, which can be observed as a dip in the scattering cross sections. Most importantly, the resonant condition is accompanied by an increased electric field enhancement, and thus electromagnetic internal energy available for interaction with light. Figure~\ref{fig:fig1}h shows the FDTD simulations of the scattering cross sections for the dual layer WS$_2$/MoS$_2$ hexagonal antenna that we designed. The anapole scattering dip is observed to shift from approximately 1300~nm to 1600~nm as the radial size of the hexagonal antenna increases from 260~nm to 360~nm. 

\begin{figure}
	\centering
	\includegraphics[width=1\linewidth]{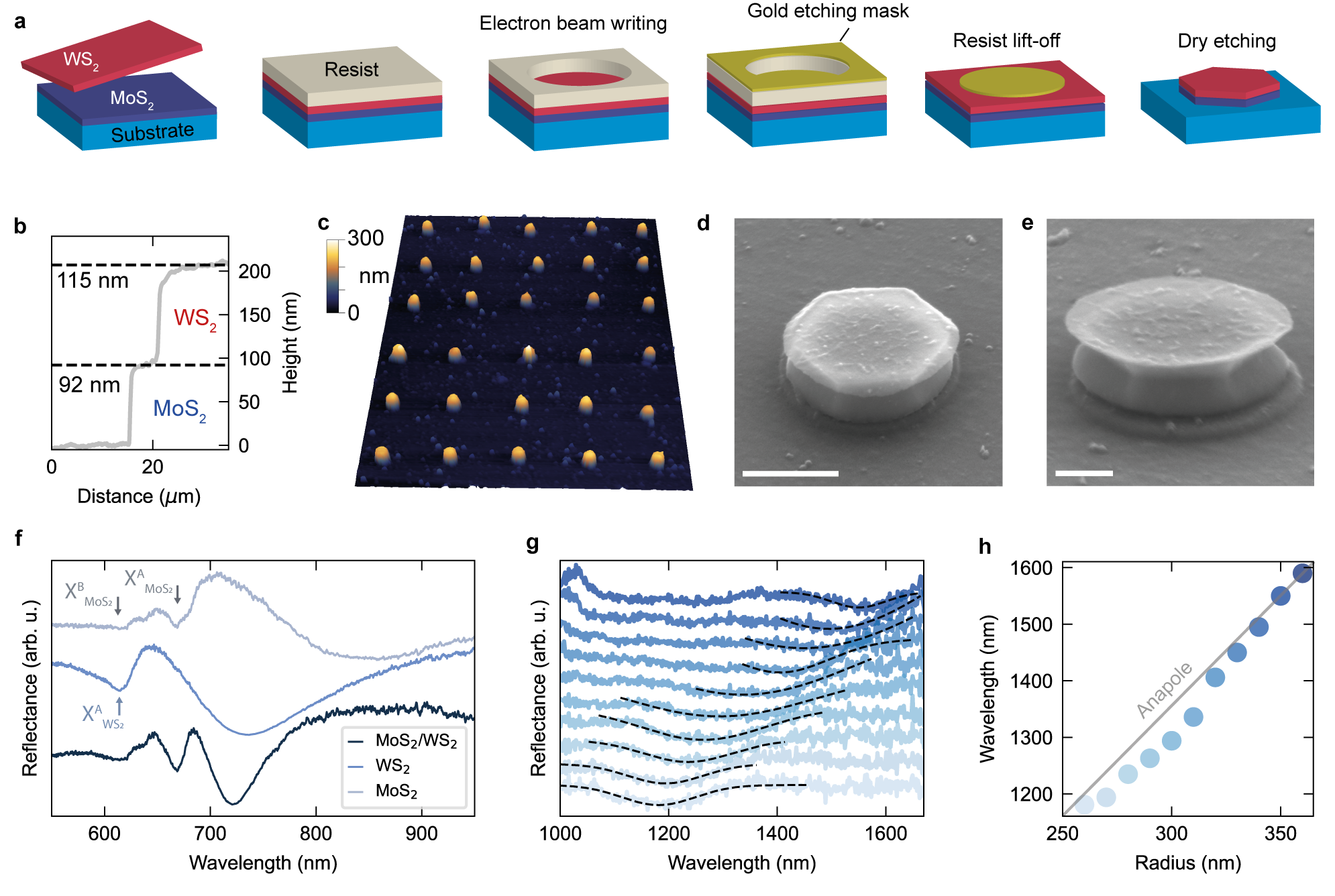}
	\caption{\textbf{Fabrication and linear optical characterization of WS$_2$/MoS$_2$ anapole nanoantennas} (a) Fabrication steps of the dual TMDC layer nanoantennas. From left to right: exfoliation and transfer of the thin TMDC films, deposition of the electron beam resist, electron beam lithography writing, gold etching mask deposition, resist lift-off and final dry etching step and mask removal (not shown). (b) Height profile of the exfoliated and stacked TMDC layers, before nanofabrication, revealing a total height of 207~nm. (c) Large area atomic force microscopy (AFM) scan of the final sample after nanofabrication. (d-e) Electron microscope images of the fabricated sample, for different tilting angles of 50 (d) and 60 (e) degrees. Scale bars: 200~nm. (f) Linear reflectance spectra of the exfoliated and unpatterned reference of MoS$_2$ and WS$_2$, with the relative A ($X^{A}$) and B ($X^{B}$) exciton energies, and that of the reference double layer unpatterned stack. (g) Reflectance of a set of WS$_2$/MoS$_2$ antennas with radius ranging from 260~nm to 360~nm. The black dashed lines are a Gaussian fit to the relative anapole spectral dip. (h) Spectral position of the anapole condition, extracted from the fit in panel (g), as a function of the radius. The grey line correspond to the linear dependence predicted from numerical simulations.}
	\label{fig:fig2}
\end{figure}

The fabrication steps of the double-layer nanoantennas are depicted in Figure~\ref{fig:fig2}a. Starting from individual TMDCs layers being mechanically exfoliated on glass substrates from commercially available bulk single crystals, we identified large area and uniform thickness multilayer with an height of approximately 100~nm. The top WS$_2$ layer is then transferred with a hot pick-up technique \cite{Purdie2018} on top of the bottom MoS$_2$ layer (Supplementary Fig.1a). 
As exfoliation procedures do not allow a fine tunining of the layer thickness, the final heterostructure exhibit an asymmetry in the relative height of the TMDC layers. Figure~\ref{fig:fig2}b shows the measured height profile of the hetero-bilayer before the nanofabrication, revealing a thickness of 92~nm (115~nm) for the MoS$_2$ (WS$_2$) layer.
After proceeding with an electron beam lithography step and deposition of a gold etching mask, the sample is exposed to a dry etching procedure where only the exposed material is left on the substrate (Supplementary Figure~1b). After fabrication, we characterized the sample via atomic force microscopy (AFM) and scanning electron microscopy (SEM). Figure~\ref{fig:fig2}c shows a large area AFM scan of an array of optical resonators after the fabrication. As further shown in Supplementary Figure~2, the height of the fabricated nanopillars is consistent with pre-fabrication data and no changes in the TMDCs thickness is observed. We then confirm the fabrication of hexagonal nanoantennas from SEM imaging, as shown in Figure~\ref{fig:fig2}d-e. For all the nanoantennas, we observe the presence of a thin ($<$5~nm) WS$_2$ film on top of the nanoresonators, a leftover of the etching process, which is not expected to modify the optical response of the nanoresonators \cite{Verre2018}. Moreover, the etching process yields tilted sidewalls, expected to blueshift the resonance, compared to simulations, due to a reduction of the resonator volume (see Supplementary Note III). However, from the etched sidewalls, we can observe the precise crystal alignment between the two TMDC layers.

We characterized the sample via visible linear reflectance spectroscopy on unpatterned reference patches of the hetero-bilayer and single TMDCs layers, as shown in Figure~\ref{fig:fig2}f. In the reflectance spectra of the MoS$_2$ and WS$_2$ layers, we identified the dominant A and B exciton resonances, also present in the unpatterned WS$_2$/MoS$_2$ heterostructure (see also Figure~\ref{fig:fig1}d). Specifically, we observed the MoS$_2$ A exciton ($X^{A}_{MoS_2}$) at approximately 700 nm, and the MoS$_2$ B exciton ($X^{B}_{MoS_2}$) along with the WS$_2$ A exciton ($X^{A}_{WS_2}$) closely resonant in energy at 630~nm. Supplementary Note IV includes the visible reflectance spectra of the fabricated nanoantennas, where these exciton signatures are also evident. Moving to the near-infrared region, we detected a dip in the reflectance of the double-layer nanostructures, indicative of the anapole state. Figure~\ref{fig:fig2}g presents the experimental reflectance data for a series of nanoantennas with varying radial sizes, along with the corresponding fit of the anapole dip. The dip's spectral position reveals the expected linear dependence of the anapole wavelength on the nanoantenna radius, ranging from 1200~nm to 1600~nm (Figure~\ref{fig:fig2}h), consistent with predictions from numerical simulations (see also Supplementary Figure 3 and 4).

\begin{figure}
	\centering
	\includegraphics[width=1\linewidth]{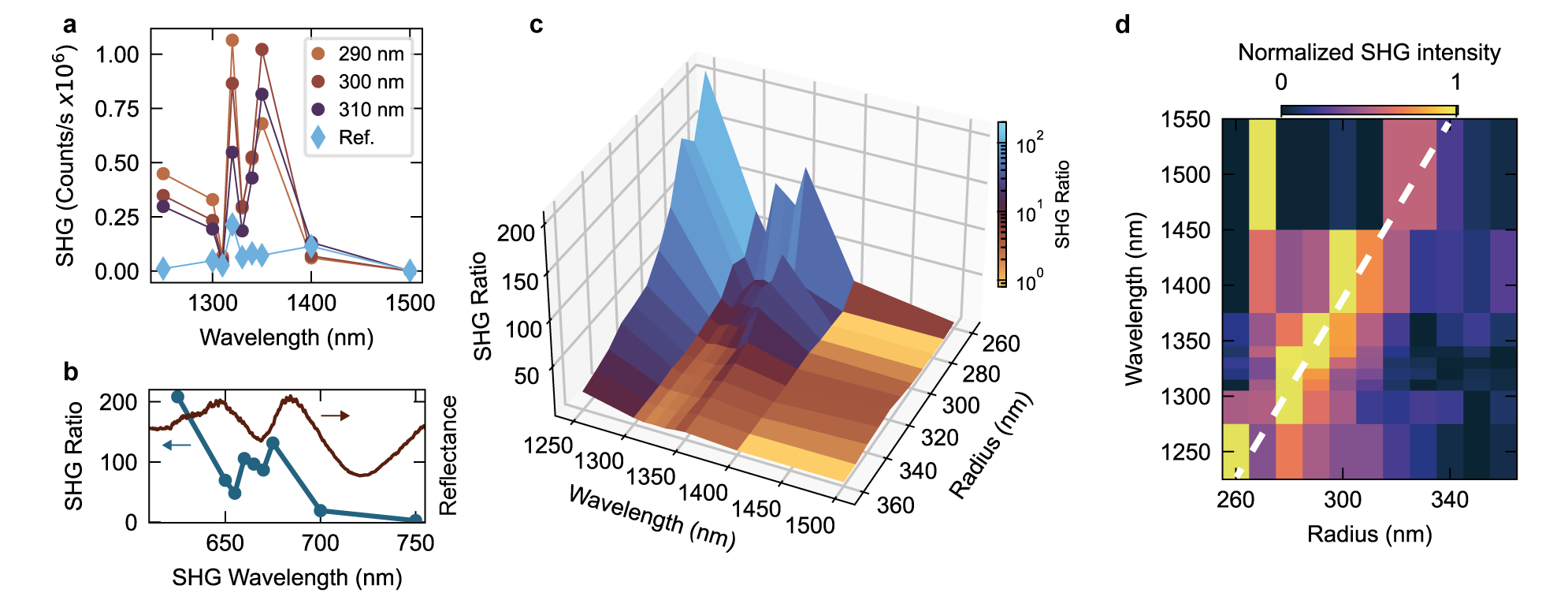}
	\caption{\textbf{Exciton and anapole driven enhancement of WS$_2$/MoS$_2$ interface second harmonic generation} (a) Raw counts of the SHG intensity as a function of the fundamental wavelength, for nanoantennas with radii of 290~nm, 300~nm and 310~nm (dot markers), and the reference sample (diamond markers). (b) Comparison between the SHG ratio for a nanoantenna with radius of 280~nm and the reflectance spectrum of the reference hetero-bilayer, where the SHG enhancement is resonant with the exciton position of the TMDC layers. (c) Three-dimensional plot of the SHG ratio, as a function of the nanoantenna radius and fundamental wavelength. (d) Normalized SHG signal for the same set of nanoantennas in panel 3c, revealing a linear dependence of the maximum SHG emission and the anapole wavelength (dashed white line).}
	\label{fig:fig3}
\end{figure}

To investigate the WS$_2$/MoS$_2$ nanoantennas in the non-linear regime, we excite nanoresonators of different radial size with fundamental wavelengths in the range from 1250~nm to 1500~nm employing a tunable optical parametric amplifier (see Methods for more details).
We observe a strong SHG enhancement in the nanostructured samples, as shown in Figure~\ref{fig:fig3}a. Here, we plot the amplitude of the SH signal intensity of nanoantennas with different radii (circular markers), compared to the reference hetero-bilayer sample (diamond markers). The strong enhancement of the SHG is closely correlated with the TMDCs exciton resonance position, as can be observed in Figure~\ref{fig:fig3}b. Here, we define the SHG enhancement ratio (see also Supplementary Note V) as the SHG intensity counts collected from the large area reference sample ($I_{ref}$), normalized over the laser spot area ($A_{laser}$), and its ratio with the SHG collected from the nanoantennas ($I_r$), normalized over the relative cross sectional area ($A_r$) extracted from SEM images.
When the fundamental laser is resonant with $X^{A}_{MoS_2}$ we observe two orders of magnitude enhancement and a peculiar double peak, replicated in both antennas and reference sample, which we ascribe to the excitonic structure of MoS$_2$. We also observe even larger values at 625~nm, most likely due to the resonant $X^{B}_{MoS_2}$ and $X^{A}_{WS_2}$ states overlapping in energy. Figure~\ref{fig:fig3}c shows the SHG enhancement ratio for the whole range of WS$_2$/MoS$_2$ nanoantennas, as a function of the radial size and fundamental excitation wavelength. In smaller radii nanoantennas, where the anapole state is also expected to be resonant with the exciton energy, we observe up to two orders of magnitude enhancement. To understand the role of the anapole field confinement on enhancing the SHG signal, we plot in Figure~\ref{fig:fig3}d the normalized SHG intensity from the same set of nanoantennas. We  observe that the increased SHG signal follows a linear dependence with the nanoantenna radius, matching the anapole wavelength extracted in Figure~\ref{fig:fig2}g. These observations demonstrate the non-trivial interplay on the SHG emission by the combined action of excitonic bulk resonances, controlled by the choice of the layered vdW material, and the anapole field confinement, tailored via the nanostructure geometry. This dual influence, material selection and geometric tailoring, highlights the complexity of the system, offering deeper insights into the mechanisms governing nonlinear optical processes in multilayered vdW nanophotonic structures.

\section{Discussion}

In summary, our work represents the first demonstration, to our knowledge, of using vdW heterostructures, specifically TMDCs, to fabricate optical nanoantennas to target nonlinear optical processes. We observed a significant and non-trivial SHG enhancement from the WS$_2$/MoS$_2$ interface, driven by the interplay of excitonic and photonic resonances of the double-layer nanoantenna. The distinct structural and optical properties of vdW materials position them as ideal candidates for future applications in nanophotonics and non-linear optics, offering new degrees of freedom in design. Moreover, the inherent anisotropies and the tunability of the twist angle between vdW crystal axes present exciting opportunities for generating nonlinear light at material interfaces which could present Moir\'{e}-induced effects, with both theoretical and practical implications. Our multilayer approach can be further extended to the broad library of vdW materials. As vdW materials enable the arbitrary stacking of different crystal structures, a deep understanding of their interfacial nonlinear properties opens to the development of engineered multilayered nanostructures and optical metamaterials optimized for enhanced light-matter interactions and non-linear optics.

\newpage

\bigskip
\noindent
\textbf{Methods}

\noindent
\textbf{Fabrication} WS$_2$ and MoS$_2$ thin films were mechanically exfoliated onto commercial silicon/silicon dioxide wafers. Flakes of suitable size and thickness were selected using optical microscopy and a profilometer (Bruker Dektak). For the fabrication of the TMDC hetero-bilayer, first the WS$_2$ flake was picked up using a polydimethylsiloxane stamp with a thin layer of poly(bisphenol A carbonate) on top. The WS$_2$ flake was subsequently brought in contact with the MoS$_2$ flake, picked up and both components were finally stamped on a fused silica substrate. The TMDC stack was then patterned into single disk structures following reference \cite{Weber2022a}.

\noindent
\textbf{Linear spectroscopy}
The linear spectroscopy setup schematics is shown in Supplementary Figure~7. We employ a tungsten lamp (Thorlabs, SLS201L) collimated by a parabolic mirror (Thorlabs, RC08FC-P01) as a white light source for both visible and NIR measurements. A polarizer (Thorlabs, GL10) and an half-waveplate (Thorlabs, AHWP10M-1600 or AHWP10M-850) controls the impinging polarization for NIR measurements. The light passes through a beam splitter (Thorlabs, BSS10R and BSN12R for visible and NIR, respectively). We use a three-axis piezo-motor stage (SmarAct) to precisely move the sample in the focal spot of the objective (Olympus, NIR 100x NA=0.85). The reflected light is collected by a parabolic mirror (Thorlabs, RC12FC-P01) which is coupled to an optical multimode fiber connected to a visible (Andor) or NIR (Ocean Optics, NIRQuest512) spectrometer. A flip mirror routes the reflected light to a CCD camera to visualize the sample position. The normalization procedure and background subtraction is performed according to \cite{Franceschini2023}, where we use the silica substrate as a reference instead of a silver mirror.

\noindent
\textbf{Nonlinear spectroscopy}
The nonlinear spectroscopy setup schematics is shown in Supplementary Figure 8. We feed a Monaco laser (Coherent) at 1035~nm with 300~fs pulses and a 500~kHz repetition rate to a commercial optical parametric amplifier (Coherent, Opera-F) to tune the output wavelength between 1200~nm and 2000~nm. The output radiation is filtered by a longpass pass filter (Thorlabs, FELH1100) to remove the residual pump and a bandpass filter to select the desired wavelength (Thorlabs, FB1XX0-12 series) with a 12~nm bandwidth. The fundamental beam power is controlled by an half-wave plate (Thorlabs, AHWP10M-1600) and a polarizer (Thorlabs, GL10). The input polarization can be changed by means of an additional half-wave plate. The fundamental laser beam is transmitted by a dichroic mirror (Thorlabs, DMLP950) and then focused by a large numerical aperture objective (Olympus, NIR 100x NA=0.85). The collected second harmonic is reflected by the dichroic mirror and filtered by a shortpass filter (Thorlabs, FESH800) before being focused (Thorlabs, LA1433) on a single photon avalanche detector (MPD, PD-50). A flip mirror and a system with a LED and a visible camera allow to image the samples. The actual laser spot size at the fundamental wavelength is determined by performing knife edge measurements.

\bigskip
\noindent
\textbf{Data Availability}

\noindent
The data that support the findings of this study are available from the corresponding authors upon reasonable request. 

\bibliography{dualanapole}

\begin{thebibliography}{47}%
\makeatletter
\providecommand \@ifxundefined [1]{%
 \@ifx{#1\undefined}
}%
\providecommand \@ifnum [1]{%
 \ifnum #1\expandafter \@firstoftwo
 \else \expandafter \@secondoftwo
 \fi
}%
\providecommand \@ifx [1]{%
 \ifx #1\expandafter \@firstoftwo
 \else \expandafter \@secondoftwo
 \fi
}%
\providecommand \natexlab [1]{#1}%
\providecommand \enquote  [1]{``#1''}%
\providecommand \bibnamefont  [1]{#1}%
\providecommand \bibfnamefont [1]{#1}%
\providecommand \citenamefont [1]{#1}%
\providecommand \href@noop [0]{\@secondoftwo}%
\providecommand \href [0]{\begingroup \@sanitize@url \@href}%
\providecommand \@href[1]{\@@startlink{#1}\@@href}%
\providecommand \@@href[1]{\endgroup#1\@@endlink}%
\providecommand \@sanitize@url [0]{\catcode `\\12\catcode `\$12\catcode
  `\&12\catcode `\#12\catcode `\^12\catcode `\_12\catcode `\%12\relax}%
\providecommand \@@startlink[1]{}%
\providecommand \@@endlink[0]{}%
\providecommand \url  [0]{\begingroup\@sanitize@url \@url }%
\providecommand \@url [1]{\endgroup\@href {#1}{\urlprefix }}%
\providecommand \urlprefix  [0]{URL }%
\providecommand \Eprint [0]{\href }%
\providecommand \doibase [0]{https://doi.org/}%
\providecommand \selectlanguage [0]{\@gobble}%
\providecommand \bibinfo  [0]{\@secondoftwo}%
\providecommand \bibfield  [0]{\@secondoftwo}%
\providecommand \translation [1]{[#1]}%
\providecommand \BibitemOpen [0]{}%
\providecommand \bibitemStop [0]{}%
\providecommand \bibitemNoStop [0]{.\EOS\space}%
\providecommand \EOS [0]{\spacefactor3000\relax}%
\providecommand \BibitemShut  [1]{\csname bibitem#1\endcsname}%
\let\auto@bib@innerbib\@empty
\bibitem [{\citenamefont {Kuznetsov}\ \emph {et~al.}(2016)\citenamefont
  {Kuznetsov}, \citenamefont {Miroshnichenko}, \citenamefont {Brongersma},
  \citenamefont {Kivshar},\ and\ \citenamefont
  {Luk’yanchuk}}]{Kuznetsov2016}%
  \BibitemOpen
  \bibfield  {author} {\bibinfo {author} {\bibfnamefont {A.~I.}\ \bibnamefont
  {Kuznetsov}}, \bibinfo {author} {\bibfnamefont {A.~E.}\ \bibnamefont
  {Miroshnichenko}}, \bibinfo {author} {\bibfnamefont {M.~L.}\ \bibnamefont
  {Brongersma}}, \bibinfo {author} {\bibfnamefont {Y.~S.}\ \bibnamefont
  {Kivshar}},\ and\ \bibinfo {author} {\bibfnamefont {B.}~\bibnamefont
  {Luk’yanchuk}},\ }\href {https://doi.org/10.1126/science.aag2472}
  {\bibfield  {journal} {\bibinfo  {journal} {Science}\ }\textbf {\bibinfo
  {volume} {354}},\ \bibinfo {pages} {aag2472} (\bibinfo {year}
  {2016})}\BibitemShut {NoStop}%
\bibitem [{\citenamefont {Cambiasso}\ \emph {et~al.}(2017)\citenamefont
  {Cambiasso}, \citenamefont {Grinblat}, \citenamefont {Li}, \citenamefont
  {Rakovich}, \citenamefont {Cortés},\ and\ \citenamefont
  {Maier}}]{Cambiasso2017a}%
  \BibitemOpen
  \bibfield  {author} {\bibinfo {author} {\bibfnamefont {J.}~\bibnamefont
  {Cambiasso}}, \bibinfo {author} {\bibfnamefont {G.}~\bibnamefont {Grinblat}},
  \bibinfo {author} {\bibfnamefont {Y.}~\bibnamefont {Li}}, \bibinfo {author}
  {\bibfnamefont {A.}~\bibnamefont {Rakovich}}, \bibinfo {author}
  {\bibfnamefont {E.}~\bibnamefont {Cort\'{e}s}},\ and\ \bibinfo {author}
  {\bibfnamefont {S.~A.}\ \bibnamefont {Maier}},\ }\href
  {https://doi.org/10.1021/acs.nanolett.6b05026} {\bibfield  {journal}
  {\bibinfo  {journal} {Nano Letters}\ }\textbf {\bibinfo {volume} {17}},\
  \bibinfo {pages} {1219} (\bibinfo {year} {2017})}\BibitemShut {NoStop}%
\bibitem [{\citenamefont {Xu}\ \emph {et~al.}(2020)\citenamefont {Xu},
  \citenamefont {Saerens}, \citenamefont {Timofeeva}, \citenamefont {Smirnova},
  \citenamefont {Volkovskaya}, \citenamefont {Lysevych}, \citenamefont
  {Camacho-Morales}, \citenamefont {Cai}, \citenamefont {Kamali}, \citenamefont
  {Huang}, \citenamefont {Karouta}, \citenamefont {Tan}, \citenamefont
  {Jagadish}, \citenamefont {Miroshnichenko}, \citenamefont {Grange},
  \citenamefont {Neshev},\ and\ \citenamefont {Rahmani}}]{Xu2020}%
  \BibitemOpen
  \bibfield  {author} {\bibinfo {author} {\bibfnamefont {L.}~\bibnamefont
  {Xu}}, \bibinfo {author} {\bibfnamefont {G.}~\bibnamefont {Saerens}},
  \bibinfo {author} {\bibfnamefont {M.}~\bibnamefont {Timofeeva}}, \bibinfo
  {author} {\bibfnamefont {D.~A.}\ \bibnamefont {Smirnova}}, \bibinfo {author}
  {\bibfnamefont {I.}~\bibnamefont {Volkovskaya}}, \bibinfo {author}
  {\bibfnamefont {M.}~\bibnamefont {Lysevych}}, \bibinfo {author}
  {\bibfnamefont {R.}~\bibnamefont {Camacho-Morales}}, \bibinfo {author}
  {\bibfnamefont {M.}~\bibnamefont {Cai}}, \bibinfo {author} {\bibfnamefont
  {K.~Z.}\ \bibnamefont {Kamali}}, \bibinfo {author} {\bibfnamefont
  {L.}~\bibnamefont {Huang}}, \bibinfo {author} {\bibfnamefont
  {F.}~\bibnamefont {Karouta}}, \bibinfo {author} {\bibfnamefont {H.~H.}\
  \bibnamefont {Tan}}, \bibinfo {author} {\bibfnamefont {C.}~\bibnamefont
  {Jagadish}}, \bibinfo {author} {\bibfnamefont {A.~E.}\ \bibnamefont
  {Miroshnichenko}}, \bibinfo {author} {\bibfnamefont {R.}~\bibnamefont
  {Grange}}, \bibinfo {author} {\bibfnamefont {D.~N.}\ \bibnamefont {Neshev}},\
  and\ \bibinfo {author} {\bibfnamefont {M.}~\bibnamefont {Rahmani}},\ }\href
  {https://doi.org/10.1021/acsnano.9b07117} {\bibfield  {journal} {\bibinfo
  {journal} {ACS Nano}\ }\textbf {\bibinfo {volume} {14}},\ \bibinfo {pages}
  {1379} (\bibinfo {year} {2020})}\BibitemShut {NoStop}%
\bibitem [{\citenamefont {Kuznetsov}\ \emph {et~al.}(2012)\citenamefont
  {Kuznetsov}, \citenamefont {Miroshnichenko}, \citenamefont {Fu},
  \citenamefont {Zhang},\ and\ \citenamefont {Luk’yanchuk}}]{Kuznetsov2012}%
  \BibitemOpen
  \bibfield  {author} {\bibinfo {author} {\bibfnamefont {A.~I.}\ \bibnamefont
  {Kuznetsov}}, \bibinfo {author} {\bibfnamefont {A.~E.}\ \bibnamefont
  {Miroshnichenko}}, \bibinfo {author} {\bibfnamefont {Y.~H.}\ \bibnamefont
  {Fu}}, \bibinfo {author} {\bibfnamefont {J.}~\bibnamefont {Zhang}},\ and\
  \bibinfo {author} {\bibfnamefont {B.}~\bibnamefont {Luk’yanchuk}},\ }\href
  {https://doi.org/10.1038/srep00492} {\bibfield  {journal} {\bibinfo
  {journal} {Scientific Reports}\ }\textbf {\bibinfo {volume} {2}},\ \bibinfo
  {pages} {492} (\bibinfo {year} {2012})}\BibitemShut {NoStop}%
\bibitem [{\citenamefont {Liu}\ and\ \citenamefont {Kivshar}(2017)}]{Liu2017a}%
  \BibitemOpen
  \bibfield  {author} {\bibinfo {author} {\bibfnamefont {W.}~\bibnamefont
  {Liu}}\ and\ \bibinfo {author} {\bibfnamefont {Y.~S.}\ \bibnamefont
  {Kivshar}},\ }\href {https://doi.org/10.1098/rsta.2016.0317} {\bibfield
  {journal} {\bibinfo  {journal} {Philosophical Transactions of the Royal
  Society A: Mathematical, Physical and Engineering Sciences}\ }\textbf
  {\bibinfo {volume} {375}},\ \bibinfo {pages} {20160317} (\bibinfo {year}
  {2017})}\BibitemShut {NoStop}%
\bibitem [{\citenamefont {Koshelev}\ \emph {et~al.}(2019)\citenamefont
  {Koshelev}, \citenamefont {Favraud}, \citenamefont {Bogdanov}, \citenamefont
  {Kivshar},\ and\ \citenamefont {Fratalocchi}}]{Koshelev2019}%
  \BibitemOpen
  \bibfield  {author} {\bibinfo {author} {\bibfnamefont {K.}~\bibnamefont
  {Koshelev}}, \bibinfo {author} {\bibfnamefont {G.}~\bibnamefont {Favraud}},
  \bibinfo {author} {\bibfnamefont {A.}~\bibnamefont {Bogdanov}}, \bibinfo
  {author} {\bibfnamefont {Y.}~\bibnamefont {Kivshar}},\ and\ \bibinfo {author}
  {\bibfnamefont {A.}~\bibnamefont {Fratalocchi}},\ }\href
  {https://doi.org/10.1515/nanoph-2019-0024} {\bibfield  {journal} {\bibinfo
  {journal} {Nanophotonics}\ }\textbf {\bibinfo {volume} {8}},\ \bibinfo
  {pages} {725} (\bibinfo {year} {2019})}\BibitemShut {NoStop}%
\bibitem [{\citenamefont {Miroshnichenko}\ \emph {et~al.}(2015)\citenamefont
  {Miroshnichenko}, \citenamefont {Evlyukhin}, \citenamefont {Yu},
  \citenamefont {Bakker}, \citenamefont {Chipouline}, \citenamefont
  {Kuznetsov}, \citenamefont {Luk’yanchuk}, \citenamefont {Chichkov},\ and\
  \citenamefont {Kivshar}}]{Miroshnichenko2015}%
  \BibitemOpen
  \bibfield  {author} {\bibinfo {author} {\bibfnamefont {A.~E.}\ \bibnamefont
  {Miroshnichenko}}, \bibinfo {author} {\bibfnamefont {A.~B.}\ \bibnamefont
  {Evlyukhin}}, \bibinfo {author} {\bibfnamefont {Y.~F.}\ \bibnamefont {Yu}},
  \bibinfo {author} {\bibfnamefont {R.~M.}\ \bibnamefont {Bakker}}, \bibinfo
  {author} {\bibfnamefont {A.}~\bibnamefont {Chipouline}}, \bibinfo {author}
  {\bibfnamefont {A.~I.}\ \bibnamefont {Kuznetsov}}, \bibinfo {author}
  {\bibfnamefont {B.}~\bibnamefont {Luk’yanchuk}}, \bibinfo {author}
  {\bibfnamefont {B.~N.}\ \bibnamefont {Chichkov}},\ and\ \bibinfo {author}
  {\bibfnamefont {Y.~S.}\ \bibnamefont {Kivshar}},\ }\href
  {https://doi.org/10.1038/ncomms9069} {\bibfield  {journal} {\bibinfo
  {journal} {Nature Communications}\ }\textbf {\bibinfo {volume} {6}},\
  \bibinfo {pages} {8069} (\bibinfo {year} {2015})}\BibitemShut {NoStop}%
\bibitem [{\citenamefont {Kang}\ \emph {et~al.}(2023)\citenamefont {Kang},
  \citenamefont {Liu}, \citenamefont {Chan},\ and\ \citenamefont
  {Xiao}}]{Kang2023}%
  \BibitemOpen
  \bibfield  {author} {\bibinfo {author} {\bibfnamefont {M.}~\bibnamefont
  {Kang}}, \bibinfo {author} {\bibfnamefont {T.}~\bibnamefont {Liu}}, \bibinfo
  {author} {\bibfnamefont {C.~T.}\ \bibnamefont {Chan}},\ and\ \bibinfo
  {author} {\bibfnamefont {M.}~\bibnamefont {Xiao}},\ }\href
  {https://doi.org/10.1038/s42254-023-00642-8} {\bibfield  {journal} {\bibinfo
  {journal} {Nature Reviews Physics}\ }\textbf {\bibinfo {volume} {5}},\
  \bibinfo {pages} {659} (\bibinfo {year} {2023})}\BibitemShut {NoStop}%
\bibitem [{\citenamefont {Carletti}\ \emph {et~al.}(2018)\citenamefont
  {Carletti}, \citenamefont {Koshelev}, \citenamefont {Angelis},\ and\
  \citenamefont {Kivshar}}]{Carletti2018}%
  \BibitemOpen
  \bibfield  {author} {\bibinfo {author} {\bibfnamefont {L.}~\bibnamefont
  {Carletti}}, \bibinfo {author} {\bibfnamefont {K.}~\bibnamefont {Koshelev}},
  \bibinfo {author} {\bibfnamefont {C.~D.}\ \bibnamefont {Angelis}},\ and\
  \bibinfo {author} {\bibfnamefont {Y.}~\bibnamefont {Kivshar}},\ }\href
  {https://doi.org/10.1103/physrevlett.121.033903} {\bibfield  {journal}
  {\bibinfo  {journal} {Physical Review Letters}\ }\textbf {\bibinfo {volume}
  {121}},\ \bibinfo {pages} {033903} (\bibinfo {year} {2018})}\BibitemShut
  {NoStop}%
\bibitem [{\citenamefont {Koshelev}\ \emph {et~al.}(2020)\citenamefont
  {Koshelev}, \citenamefont {Kruk}, \citenamefont {Melik-Gaykazyan},
  \citenamefont {Choi}, \citenamefont {Bogdanov}, \citenamefont {Park},\ and\
  \citenamefont {Kivshar}}]{Koshelev2020}%
  \BibitemOpen
  \bibfield  {author} {\bibinfo {author} {\bibfnamefont {K.}~\bibnamefont
  {Koshelev}}, \bibinfo {author} {\bibfnamefont {S.}~\bibnamefont {Kruk}},
  \bibinfo {author} {\bibfnamefont {E.}~\bibnamefont {Melik-Gaykazyan}},
  \bibinfo {author} {\bibfnamefont {J.-H.}\ \bibnamefont {Choi}}, \bibinfo
  {author} {\bibfnamefont {A.}~\bibnamefont {Bogdanov}}, \bibinfo {author}
  {\bibfnamefont {H.-G.}\ \bibnamefont {Park}},\ and\ \bibinfo {author}
  {\bibfnamefont {Y.}~\bibnamefont {Kivshar}},\ }\href
  {https://doi.org/10.1126/science.aaz3985} {\bibfield  {journal} {\bibinfo
  {journal} {Science}\ }\textbf {\bibinfo {volume} {367}},\ \bibinfo {pages}
  {288} (\bibinfo {year} {2020})}\BibitemShut {NoStop}%
\bibitem [{\citenamefont {Meng}\ \emph {et~al.}(2023)\citenamefont {Meng},
  \citenamefont {Feng}, \citenamefont {Han}, \citenamefont {Xu}, \citenamefont
  {Mao}, \citenamefont {Zhang}, \citenamefont {Kim}, \citenamefont {Roh},
  \citenamefont {Zhao}, \citenamefont {Kim}, \citenamefont {Yang},
  \citenamefont {Lee}, \citenamefont {Yang}, \citenamefont {Qiu},\ and\
  \citenamefont {Bae}}]{Meng2023}%
  \BibitemOpen
  \bibfield  {author} {\bibinfo {author} {\bibfnamefont {Y.}~\bibnamefont
  {Meng}}, \bibinfo {author} {\bibfnamefont {J.}~\bibnamefont {Feng}}, \bibinfo
  {author} {\bibfnamefont {S.}~\bibnamefont {Han}}, \bibinfo {author}
  {\bibfnamefont {Z.}~\bibnamefont {Xu}}, \bibinfo {author} {\bibfnamefont
  {W.}~\bibnamefont {Mao}}, \bibinfo {author} {\bibfnamefont {T.}~\bibnamefont
  {Zhang}}, \bibinfo {author} {\bibfnamefont {J.~S.}\ \bibnamefont {Kim}},
  \bibinfo {author} {\bibfnamefont {I.}~\bibnamefont {Roh}}, \bibinfo {author}
  {\bibfnamefont {Y.}~\bibnamefont {Zhao}}, \bibinfo {author} {\bibfnamefont
  {D.-H.}\ \bibnamefont {Kim}}, \bibinfo {author} {\bibfnamefont
  {Y.}~\bibnamefont {Yang}}, \bibinfo {author} {\bibfnamefont {J.-W.}\
  \bibnamefont {Lee}}, \bibinfo {author} {\bibfnamefont {L.}~\bibnamefont
  {Yang}}, \bibinfo {author} {\bibfnamefont {C.-W.}\ \bibnamefont {Qiu}},\ and\
  \bibinfo {author} {\bibfnamefont {S.-H.}\ \bibnamefont {Bae}},\ }\href
  {https://doi.org/10.1038/s41578-023-00558-w} {\bibfield  {journal} {\bibinfo
  {journal} {Nature Reviews Materials}\ }\textbf {\bibinfo {volume} {8}},\
  \bibinfo {pages} {498} (\bibinfo {year} {2023})}\BibitemShut {NoStop}%
\bibitem [{\citenamefont {Trovatello}\ \emph {et~al.}(2024)\citenamefont
  {Trovatello}, \citenamefont {Marini}, \citenamefont {Cotrufo}, \citenamefont
  {Alù}, \citenamefont {Schuck},\ and\ \citenamefont
  {Cerullo}}]{Trovatello2024}%
  \BibitemOpen
  \bibfield  {author} {\bibinfo {author} {\bibfnamefont {C.}~\bibnamefont
  {Trovatello}}, \bibinfo {author} {\bibfnamefont {A.}~\bibnamefont {Marini}},
  \bibinfo {author} {\bibfnamefont {M.}~\bibnamefont {Cotrufo}}, \bibinfo
  {author} {\bibfnamefont {A.}~\bibnamefont {Al\'{u}}}, \bibinfo {author}
  {\bibfnamefont {P.~J.}\ \bibnamefont {Schuck}},\ and\ \bibinfo {author}
  {\bibfnamefont {G.}~\bibnamefont {Cerullo}},\ }\href
  {https://doi.org/10.1021/acsphotonics.4c00521} {\bibfield  {journal}
  {\bibinfo  {journal} {ACS Photonics}\ }\textbf {\bibinfo {volume} {11}},\
  \bibinfo {pages} {2860} (\bibinfo {year} {2024})}\BibitemShut {NoStop}%
\bibitem [{\citenamefont {Khurgin}(2022)}]{Khurgin2022}%
  \BibitemOpen
  \bibfield  {author} {\bibinfo {author} {\bibfnamefont {J.~B.}\ \bibnamefont
  {Khurgin}},\ }\href {https://doi.org/10.1021/acsphotonics.1c01834} {\bibfield
   {journal} {\bibinfo  {journal} {ACS Photonics}\ }\textbf {\bibinfo {volume}
  {9}},\ \bibinfo {pages} {743} (\bibinfo {year} {2022})}\BibitemShut {NoStop}%
\bibitem [{\citenamefont {Vyshnevyy}\ \emph {et~al.}(2023)\citenamefont
  {Vyshnevyy}, \citenamefont {Ermolaev}, \citenamefont {Grudinin},
  \citenamefont {Voronin}, \citenamefont {Kharichkin}, \citenamefont {Mazitov},
  \citenamefont {Kruglov}, \citenamefont {Yakubovsky}, \citenamefont {Mishra},
  \citenamefont {Kirtaev}, \citenamefont {Arsenin}, \citenamefont {Novoselov},
  \citenamefont {Martin-Moreno},\ and\ \citenamefont {Volkov}}]{Vyshnevyy2023}%
  \BibitemOpen
  \bibfield  {author} {\bibinfo {author} {\bibfnamefont {A.~A.}\ \bibnamefont
  {Vyshnevyy}}, \bibinfo {author} {\bibfnamefont {G.~A.}\ \bibnamefont
  {Ermolaev}}, \bibinfo {author} {\bibfnamefont {D.~V.}\ \bibnamefont
  {Grudinin}}, \bibinfo {author} {\bibfnamefont {K.~V.}\ \bibnamefont
  {Voronin}}, \bibinfo {author} {\bibfnamefont {I.}~\bibnamefont {Kharichkin}},
  \bibinfo {author} {\bibfnamefont {A.}~\bibnamefont {Mazitov}}, \bibinfo
  {author} {\bibfnamefont {I.~A.}\ \bibnamefont {Kruglov}}, \bibinfo {author}
  {\bibfnamefont {D.~I.}\ \bibnamefont {Yakubovsky}}, \bibinfo {author}
  {\bibfnamefont {P.}~\bibnamefont {Mishra}}, \bibinfo {author} {\bibfnamefont
  {R.~V.}\ \bibnamefont {Kirtaev}}, \bibinfo {author} {\bibfnamefont {A.~V.}\
  \bibnamefont {Arsenin}}, \bibinfo {author} {\bibfnamefont {K.~S.}\
  \bibnamefont {Novoselov}}, \bibinfo {author} {\bibfnamefont {L.}~\bibnamefont
  {Martin-Moreno}},\ and\ \bibinfo {author} {\bibfnamefont {V.~S.}\
  \bibnamefont {Volkov}},\ }\href
  {https://doi.org/10.1021/acs.nanolett.3c02051} {\bibfield  {journal}
  {\bibinfo  {journal} {Nano Letters}\ }\textbf {\bibinfo {volume} {23}},\
  \bibinfo {pages} {8057} (\bibinfo {year} {2023})}\BibitemShut {NoStop}%
\bibitem [{\citenamefont {Novoselov}\ \emph {et~al.}(2016)\citenamefont
  {Novoselov}, \citenamefont {Mishchenko}, \citenamefont {Carvalho},\ and\
  \citenamefont {Neto}}]{Novoselov2016}%
  \BibitemOpen
  \bibfield  {author} {\bibinfo {author} {\bibfnamefont {K.~S.}\ \bibnamefont
  {Novoselov}}, \bibinfo {author} {\bibfnamefont {A.}~\bibnamefont
  {Mishchenko}}, \bibinfo {author} {\bibfnamefont {A.}~\bibnamefont
  {Carvalho}},\ and\ \bibinfo {author} {\bibfnamefont {A.~H.~C.}\ \bibnamefont
  {Neto}},\ }\href {https://doi.org/10.1126/science.aac9439} {\bibfield
  {journal} {\bibinfo  {journal} {Science}\ }\textbf {\bibinfo {volume}
  {353}},\ \bibinfo {pages} {aac9439} (\bibinfo {year} {2016})}\BibitemShut
  {NoStop}%
\bibitem [{\citenamefont {Autere}\ \emph {et~al.}(2018)\citenamefont {Autere},
  \citenamefont {Jussila}, \citenamefont {Dai}, \citenamefont {Wang},
  \citenamefont {Lipsanen},\ and\ \citenamefont {Sun}}]{Autere2018}%
  \BibitemOpen
  \bibfield  {author} {\bibinfo {author} {\bibfnamefont {A.}~\bibnamefont
  {Autere}}, \bibinfo {author} {\bibfnamefont {H.}~\bibnamefont {Jussila}},
  \bibinfo {author} {\bibfnamefont {Y.}~\bibnamefont {Dai}}, \bibinfo {author}
  {\bibfnamefont {Y.}~\bibnamefont {Wang}}, \bibinfo {author} {\bibfnamefont
  {H.}~\bibnamefont {Lipsanen}},\ and\ \bibinfo {author} {\bibfnamefont
  {Z.}~\bibnamefont {Sun}},\ }\href {https://doi.org/10.1002/adma.201705963}
  {\bibfield  {journal} {\bibinfo  {journal} {Advanced Materials}\ }\textbf
  {\bibinfo {volume} {30}},\ \bibinfo {pages} {1705963} (\bibinfo {year}
  {2018})}\BibitemShut {NoStop}%
\bibitem [{\citenamefont {Mueller}\ and\ \citenamefont
  {Malic}(2018)}]{Mueller2018}%
  \BibitemOpen
  \bibfield  {author} {\bibinfo {author} {\bibfnamefont {T.}~\bibnamefont
  {Mueller}}\ and\ \bibinfo {author} {\bibfnamefont {E.}~\bibnamefont
  {Malic}},\ }\href {https://doi.org/10.1038/s41699-018-0074-2} {\bibfield
  {journal} {\bibinfo  {journal} {npj 2D Materials and Applications}\ }\textbf
  {\bibinfo {volume} {2}},\ \bibinfo {pages} {29} (\bibinfo {year}
  {2018})}\BibitemShut {NoStop}%
\bibitem [{\citenamefont {Malard}\ \emph {et~al.}(2013)\citenamefont {Malard},
  \citenamefont {Alencar}, \citenamefont {Barboza}, \citenamefont {Mak},\ and\
  \citenamefont {Paula}}]{Malard2013}%
  \BibitemOpen
  \bibfield  {author} {\bibinfo {author} {\bibfnamefont {L.~M.}\ \bibnamefont
  {Malard}}, \bibinfo {author} {\bibfnamefont {T.~V.}\ \bibnamefont {Alencar}},
  \bibinfo {author} {\bibfnamefont {A.~P.~M.}\ \bibnamefont {Barboza}},
  \bibinfo {author} {\bibfnamefont {K.~F.}\ \bibnamefont {Mak}},\ and\ \bibinfo
  {author} {\bibfnamefont {A.~M.~d.}\ \bibnamefont {Paula}},\ }\href
  {https://doi.org/10.1103/physrevb.87.201401} {\bibfield  {journal} {\bibinfo
  {journal} {Physical Review B}\ }\textbf {\bibinfo {volume} {87}},\ \bibinfo
  {pages} {201401} (\bibinfo {year} {2013})}\BibitemShut {NoStop}%
\bibitem [{\citenamefont {Mennel}\ \emph {et~al.}(2018)\citenamefont {Mennel},
  \citenamefont {Furchi}, \citenamefont {Wachter}, \citenamefont {Paur},
  \citenamefont {Polyushkin},\ and\ \citenamefont {Mueller}}]{Mennel2018}%
  \BibitemOpen
  \bibfield  {author} {\bibinfo {author} {\bibfnamefont {L.}~\bibnamefont
  {Mennel}}, \bibinfo {author} {\bibfnamefont {M.~M.}\ \bibnamefont {Furchi}},
  \bibinfo {author} {\bibfnamefont {S.}~\bibnamefont {Wachter}}, \bibinfo
  {author} {\bibfnamefont {M.}~\bibnamefont {Paur}}, \bibinfo {author}
  {\bibfnamefont {D.~K.}\ \bibnamefont {Polyushkin}},\ and\ \bibinfo {author}
  {\bibfnamefont {T.}~\bibnamefont {Mueller}},\ }\href
  {https://doi.org/10.1038/s41467-018-02830-y} {\bibfield  {journal} {\bibinfo
  {journal} {Nature Communications}\ }\textbf {\bibinfo {volume} {9}},\
  \bibinfo {pages} {516} (\bibinfo {year} {2018})}\BibitemShut {NoStop}%
\bibitem [{\citenamefont {Lin}\ \emph {et~al.}(2019)\citenamefont {Lin},
  \citenamefont {Bange},\ and\ \citenamefont {Lupton}}]{Lin2018}%
  \BibitemOpen
  \bibfield  {author} {\bibinfo {author} {\bibfnamefont {K.-Q.}\ \bibnamefont
  {Lin}}, \bibinfo {author} {\bibfnamefont {S.}~\bibnamefont {Bange}},\ and\
  \bibinfo {author} {\bibfnamefont {J.~M.}\ \bibnamefont {Lupton}},\ }\href
  {https://doi.org/10.1038/s41567-018-0384-5} {\bibfield  {journal} {\bibinfo
  {journal} {Nature Physics}\ }\textbf {\bibinfo {volume} {15}},\ \bibinfo
  {pages} {242} (\bibinfo {year} {2019})}\BibitemShut {NoStop}%
\bibitem [{\citenamefont {Trovatello}\ \emph {et~al.}(2021)\citenamefont
  {Trovatello}, \citenamefont {Marini}, \citenamefont {Xu}, \citenamefont
  {Lee}, \citenamefont {Liu}, \citenamefont {Curreli}, \citenamefont {Manzoni},
  \citenamefont {Conte}, \citenamefont {Yao}, \citenamefont {Ciattoni},
  \citenamefont {Hone}, \citenamefont {Zhu}, \citenamefont {Schuck},\ and\
  \citenamefont {Cerullo}}]{Trovatello2020b}%
  \BibitemOpen
  \bibfield  {author} {\bibinfo {author} {\bibfnamefont {C.}~\bibnamefont
  {Trovatello}}, \bibinfo {author} {\bibfnamefont {A.}~\bibnamefont {Marini}},
  \bibinfo {author} {\bibfnamefont {X.}~\bibnamefont {Xu}}, \bibinfo {author}
  {\bibfnamefont {C.}~\bibnamefont {Lee}}, \bibinfo {author} {\bibfnamefont
  {F.}~\bibnamefont {Liu}}, \bibinfo {author} {\bibfnamefont {N.}~\bibnamefont
  {Curreli}}, \bibinfo {author} {\bibfnamefont {C.}~\bibnamefont {Manzoni}},
  \bibinfo {author} {\bibfnamefont {S.~D.}\ \bibnamefont {Conte}}, \bibinfo
  {author} {\bibfnamefont {K.}~\bibnamefont {Yao}}, \bibinfo {author}
  {\bibfnamefont {A.}~\bibnamefont {Ciattoni}}, \bibinfo {author}
  {\bibfnamefont {J.}~\bibnamefont {Hone}}, \bibinfo {author} {\bibfnamefont
  {X.}~\bibnamefont {Zhu}}, \bibinfo {author} {\bibfnamefont {P.~J.}\
  \bibnamefont {Schuck}},\ and\ \bibinfo {author} {\bibfnamefont
  {G.}~\bibnamefont {Cerullo}},\ }\href
  {https://doi.org/10.1038/s41566-020-00728-0} {\bibfield  {journal} {\bibinfo
  {journal} {Nature Photonics}\ }\textbf {\bibinfo {volume} {15}},\ \bibinfo
  {pages} {6} (\bibinfo {year} {2021})}\BibitemShut {NoStop}%
\bibitem [{\citenamefont {Klimmer}\ \emph {et~al.}(2021)\citenamefont
  {Klimmer}, \citenamefont {Ghaebi}, \citenamefont {Gan}, \citenamefont
  {George}, \citenamefont {Turchanin}, \citenamefont {Cerullo},\ and\
  \citenamefont {Soavi}}]{Klimmer2021}%
  \BibitemOpen
  \bibfield  {author} {\bibinfo {author} {\bibfnamefont {S.}~\bibnamefont
  {Klimmer}}, \bibinfo {author} {\bibfnamefont {O.}~\bibnamefont {Ghaebi}},
  \bibinfo {author} {\bibfnamefont {Z.}~\bibnamefont {Gan}}, \bibinfo {author}
  {\bibfnamefont {A.}~\bibnamefont {George}}, \bibinfo {author} {\bibfnamefont
  {A.}~\bibnamefont {Turchanin}}, \bibinfo {author} {\bibfnamefont
  {G.}~\bibnamefont {Cerullo}},\ and\ \bibinfo {author} {\bibfnamefont
  {G.}~\bibnamefont {Soavi}},\ }\href
  {https://doi.org/10.1038/s41566-021-00859-y} {\bibfield  {journal} {\bibinfo
  {journal} {Nature Photonics}\ }\textbf {\bibinfo {volume} {15}},\ \bibinfo
  {pages} {837} (\bibinfo {year} {2021})}\BibitemShut {NoStop}%
\bibitem [{\citenamefont {Lin}\ \emph {et~al.}(2022)\citenamefont {Lin},
  \citenamefont {Zhang}, \citenamefont {Zhang}, \citenamefont {Lin},
  \citenamefont {Wen}, \citenamefont {Liang}, \citenamefont {Fu}, \citenamefont
  {Lau}, \citenamefont {Ma}, \citenamefont {Qiu},\ and\ \citenamefont
  {Jia}}]{Lin2022}%
  \BibitemOpen
  \bibfield  {author} {\bibinfo {author} {\bibfnamefont {H.}~\bibnamefont
  {Lin}}, \bibinfo {author} {\bibfnamefont {Z.}~\bibnamefont {Zhang}}, \bibinfo
  {author} {\bibfnamefont {H.}~\bibnamefont {Zhang}}, \bibinfo {author}
  {\bibfnamefont {K.-T.}\ \bibnamefont {Lin}}, \bibinfo {author} {\bibfnamefont
  {X.}~\bibnamefont {Wen}}, \bibinfo {author} {\bibfnamefont {Y.}~\bibnamefont
  {Liang}}, \bibinfo {author} {\bibfnamefont {Y.}~\bibnamefont {Fu}}, \bibinfo
  {author} {\bibfnamefont {A.~K.~T.}\ \bibnamefont {Lau}}, \bibinfo {author}
  {\bibfnamefont {T.}~\bibnamefont {Ma}}, \bibinfo {author} {\bibfnamefont
  {C.-W.}\ \bibnamefont {Qiu}},\ and\ \bibinfo {author} {\bibfnamefont
  {B.}~\bibnamefont {Jia}},\ }\href
  {https://doi.org/10.1021/acs.chemrev.2c00048} {\bibfield  {journal} {\bibinfo
   {journal} {Chemical Reviews}\ }\textbf {\bibinfo {volume} {122}},\ \bibinfo
  {pages} {15204} (\bibinfo {year} {2022})}\BibitemShut {NoStop}%
\bibitem [{\citenamefont {Munkhbat}\ \emph {et~al.}(2023)\citenamefont
  {Munkhbat}, \citenamefont {Küçüköz}, \citenamefont {Baranov},
  \citenamefont {Antosiewicz},\ and\ \citenamefont {Shegai}}]{Munkhbat2022b}%
  \BibitemOpen
  \bibfield  {author} {\bibinfo {author} {\bibfnamefont {B.}~\bibnamefont
  {Munkhbat}}, \bibinfo {author} {\bibfnamefont {B.}~\bibnamefont
  {Küçüköz}}, \bibinfo {author} {\bibfnamefont {D.~G.}\ \bibnamefont
  {Baranov}}, \bibinfo {author} {\bibfnamefont {T.~J.}\ \bibnamefont
  {Antosiewicz}},\ and\ \bibinfo {author} {\bibfnamefont {T.~O.}\ \bibnamefont
  {Shegai}},\ }\href {https://doi.org/10.1002/lpor.202200057} {\bibfield
  {journal} {\bibinfo  {journal} {Laser \& Photonics Reviews}\ }\textbf
  {\bibinfo {volume} {17}},\ \bibinfo {pages} {2200057} (\bibinfo {year}
  {2023})}\BibitemShut {NoStop}%
\bibitem [{\citenamefont {Zotev}\ \emph {et~al.}(2023)\citenamefont {Zotev},
  \citenamefont {Wang}, \citenamefont {Andres‐Penares}, \citenamefont
  {Severs‐Millard}, \citenamefont {Randerson}, \citenamefont {Hu},
  \citenamefont {Sortino}, \citenamefont {Louca}, \citenamefont
  {Brotons‐Gisbert}, \citenamefont {Huq}, \citenamefont {Vezzoli},
  \citenamefont {Sapienza}, \citenamefont {Krauss}, \citenamefont {Gerardot},\
  and\ \citenamefont {Tartakovskii}}]{Zotev2022b}%
  \BibitemOpen
  \bibfield  {author} {\bibinfo {author} {\bibfnamefont {P.~G.}\ \bibnamefont
  {Zotev}}, \bibinfo {author} {\bibfnamefont {Y.}~\bibnamefont {Wang}},
  \bibinfo {author} {\bibfnamefont {D.}~\bibnamefont {Andres‐Penares}},
  \bibinfo {author} {\bibfnamefont {T.}~\bibnamefont {Severs‐Millard}},
  \bibinfo {author} {\bibfnamefont {S.}~\bibnamefont {Randerson}}, \bibinfo
  {author} {\bibfnamefont {X.}~\bibnamefont {Hu}}, \bibinfo {author}
  {\bibfnamefont {L.}~\bibnamefont {Sortino}}, \bibinfo {author} {\bibfnamefont
  {C.}~\bibnamefont {Louca}}, \bibinfo {author} {\bibfnamefont
  {M.}~\bibnamefont {Brotons‐Gisbert}}, \bibinfo {author} {\bibfnamefont
  {T.}~\bibnamefont {Huq}}, \bibinfo {author} {\bibfnamefont {S.}~\bibnamefont
  {Vezzoli}}, \bibinfo {author} {\bibfnamefont {R.}~\bibnamefont {Sapienza}},
  \bibinfo {author} {\bibfnamefont {T.~F.}\ \bibnamefont {Krauss}}, \bibinfo
  {author} {\bibfnamefont {B.~D.}\ \bibnamefont {Gerardot}},\ and\ \bibinfo
  {author} {\bibfnamefont {A.~I.}\ \bibnamefont {Tartakovskii}},\ }\href
  {https://doi.org/10.1002/lpor.202200957} {\bibfield  {journal} {\bibinfo
  {journal} {Laser \& Photonics Reviews}\ }\textbf {\bibinfo {volume} {17}},\
  \bibinfo {pages} {2200957} (\bibinfo {year} {2023})}\BibitemShut {NoStop}%
\bibitem [{\citenamefont {Ermolaev}\ \emph {et~al.}(2021)\citenamefont
  {Ermolaev}, \citenamefont {Grudinin}, \citenamefont {Stebunov}, \citenamefont
  {Voronin}, \citenamefont {Kravets}, \citenamefont {Duan}, \citenamefont
  {Mazitov}, \citenamefont {Tselikov}, \citenamefont {Bylinkin}, \citenamefont
  {Yakubovsky}, \citenamefont {Novikov}, \citenamefont {Baranov}, \citenamefont
  {Nikitin}, \citenamefont {Kruglov}, \citenamefont {Shegai}, \citenamefont
  {Alonso-González}, \citenamefont {Grigorenko}, \citenamefont {Arsenin},
  \citenamefont {Novoselov},\ and\ \citenamefont {Volkov}}]{Ermolaev2020}%
  \BibitemOpen
  \bibfield  {author} {\bibinfo {author} {\bibfnamefont {G.~A.}\ \bibnamefont
  {Ermolaev}}, \bibinfo {author} {\bibfnamefont {D.~V.}\ \bibnamefont
  {Grudinin}}, \bibinfo {author} {\bibfnamefont {Y.~V.}\ \bibnamefont
  {Stebunov}}, \bibinfo {author} {\bibfnamefont {K.~V.}\ \bibnamefont
  {Voronin}}, \bibinfo {author} {\bibfnamefont {V.~G.}\ \bibnamefont
  {Kravets}}, \bibinfo {author} {\bibfnamefont {J.}~\bibnamefont {Duan}},
  \bibinfo {author} {\bibfnamefont {A.~B.}\ \bibnamefont {Mazitov}}, \bibinfo
  {author} {\bibfnamefont {G.~I.}\ \bibnamefont {Tselikov}}, \bibinfo {author}
  {\bibfnamefont {A.}~\bibnamefont {Bylinkin}}, \bibinfo {author}
  {\bibfnamefont {D.~I.}\ \bibnamefont {Yakubovsky}}, \bibinfo {author}
  {\bibfnamefont {S.~M.}\ \bibnamefont {Novikov}}, \bibinfo {author}
  {\bibfnamefont {D.~G.}\ \bibnamefont {Baranov}}, \bibinfo {author}
  {\bibfnamefont {A.~Y.}\ \bibnamefont {Nikitin}}, \bibinfo {author}
  {\bibfnamefont {I.~A.}\ \bibnamefont {Kruglov}}, \bibinfo {author}
  {\bibfnamefont {T.}~\bibnamefont {Shegai}}, \bibinfo {author} {\bibfnamefont
  {P.}~\bibnamefont {Alonso-González}}, \bibinfo {author} {\bibfnamefont
  {A.~N.}\ \bibnamefont {Grigorenko}}, \bibinfo {author} {\bibfnamefont
  {A.~V.}\ \bibnamefont {Arsenin}}, \bibinfo {author} {\bibfnamefont {K.~S.}\
  \bibnamefont {Novoselov}},\ and\ \bibinfo {author} {\bibfnamefont {V.~S.}\
  \bibnamefont {Volkov}},\ }\href {https://doi.org/10.1038/s41467-021-21139-x}
  {\bibfield  {journal} {\bibinfo  {journal} {Nature Communications}\ }\textbf
  {\bibinfo {volume} {12}},\ \bibinfo {pages} {854} (\bibinfo {year}
  {2021})}\BibitemShut {NoStop}%
\bibitem [{\citenamefont {Munkhbat}\ \emph {et~al.}(2022)\citenamefont
  {Munkhbat}, \citenamefont {Wróbel}, \citenamefont {Antosiewicz},\ and\
  \citenamefont {Shegai}}]{Munkhbat2022}%
  \BibitemOpen
  \bibfield  {author} {\bibinfo {author} {\bibfnamefont {B.}~\bibnamefont
  {Munkhbat}}, \bibinfo {author} {\bibfnamefont {P.}~\bibnamefont {Wr\'{o}bel}},
  \bibinfo {author} {\bibfnamefont {T.~J.}\ \bibnamefont {Antosiewicz}},\ and\
  \bibinfo {author} {\bibfnamefont {T.~O.}\ \bibnamefont {Shegai}},\ }\href
  {https://doi.org/10.1021/acsphotonics.2c00433} {\bibfield  {journal}
  {\bibinfo  {journal} {ACS Photonics}\ }\textbf {\bibinfo {volume} {9}},\
  \bibinfo {pages} {2398} (\bibinfo {year} {2022})}\BibitemShut {NoStop}%
\bibitem [{\citenamefont {Zotev}\ \emph {et~al.}(2022)\citenamefont {Zotev},
  \citenamefont {Wang}, \citenamefont {Sortino}, \citenamefont {Millard},
  \citenamefont {Mullin}, \citenamefont {Conteduca}, \citenamefont {Shagar},
  \citenamefont {Genco}, \citenamefont {Hobbs}, \citenamefont {Krauss},\ and\
  \citenamefont {Tartakovskii}}]{Zotev2022}%
  \BibitemOpen
  \bibfield  {author} {\bibinfo {author} {\bibfnamefont {P.~G.}\ \bibnamefont
  {Zotev}}, \bibinfo {author} {\bibfnamefont {Y.}~\bibnamefont {Wang}},
  \bibinfo {author} {\bibfnamefont {L.}~\bibnamefont {Sortino}}, \bibinfo
  {author} {\bibfnamefont {T.~S.}\ \bibnamefont {Millard}}, \bibinfo {author}
  {\bibfnamefont {N.}~\bibnamefont {Mullin}}, \bibinfo {author} {\bibfnamefont
  {D.}~\bibnamefont {Conteduca}}, \bibinfo {author} {\bibfnamefont
  {M.}~\bibnamefont {Shagar}}, \bibinfo {author} {\bibfnamefont
  {A.}~\bibnamefont {Genco}}, \bibinfo {author} {\bibfnamefont {J.~K.}\
  \bibnamefont {Hobbs}}, \bibinfo {author} {\bibfnamefont {T.~F.}\ \bibnamefont
  {Krauss}},\ and\ \bibinfo {author} {\bibfnamefont {A.~I.}\ \bibnamefont
  {Tartakovskii}},\ }\href {https://doi.org/10.1021/acsnano.2c00802} {\bibfield
   {journal} {\bibinfo  {journal} {ACS Nano}\ }\textbf {\bibinfo {volume}
  {16}},\ \bibinfo {pages} {6493} (\bibinfo {year} {2022})}\BibitemShut
  {NoStop}%
\bibitem [{\citenamefont {Yao}\ \emph {et~al.}(2021)\citenamefont {Yao},
  \citenamefont {Finney}, \citenamefont {Zhang}, \citenamefont {Moore},
  \citenamefont {Xian}, \citenamefont {Tancogne-Dejean}, \citenamefont {Liu},
  \citenamefont {Ardelean}, \citenamefont {Xu}, \citenamefont {Halbertal},
  \citenamefont {Watanabe}, \citenamefont {Taniguchi}, \citenamefont {Ochoa},
  \citenamefont {Asenjo-Garcia}, \citenamefont {Zhu}, \citenamefont {Basov},
  \citenamefont {Rubio}, \citenamefont {Dean}, \citenamefont {Hone},\ and\
  \citenamefont {Schuck}}]{Yao2021}%
  \BibitemOpen
  \bibfield  {author} {\bibinfo {author} {\bibfnamefont {K.}~\bibnamefont
  {Yao}}, \bibinfo {author} {\bibfnamefont {N.~R.}\ \bibnamefont {Finney}},
  \bibinfo {author} {\bibfnamefont {J.}~\bibnamefont {Zhang}}, \bibinfo
  {author} {\bibfnamefont {S.~L.}\ \bibnamefont {Moore}}, \bibinfo {author}
  {\bibfnamefont {L.}~\bibnamefont {Xian}}, \bibinfo {author} {\bibfnamefont
  {N.}~\bibnamefont {Tancogne-Dejean}}, \bibinfo {author} {\bibfnamefont
  {F.}~\bibnamefont {Liu}}, \bibinfo {author} {\bibfnamefont {J.}~\bibnamefont
  {Ardelean}}, \bibinfo {author} {\bibfnamefont {X.}~\bibnamefont {Xu}},
  \bibinfo {author} {\bibfnamefont {D.}~\bibnamefont {Halbertal}}, \bibinfo
  {author} {\bibfnamefont {K.}~\bibnamefont {Watanabe}}, \bibinfo {author}
  {\bibfnamefont {T.}~\bibnamefont {Taniguchi}}, \bibinfo {author}
  {\bibfnamefont {H.}~\bibnamefont {Ochoa}}, \bibinfo {author} {\bibfnamefont
  {A.}~\bibnamefont {Asenjo-Garcia}}, \bibinfo {author} {\bibfnamefont
  {X.}~\bibnamefont {Zhu}}, \bibinfo {author} {\bibfnamefont {D.~N.}\
  \bibnamefont {Basov}}, \bibinfo {author} {\bibfnamefont {A.}~\bibnamefont
  {Rubio}}, \bibinfo {author} {\bibfnamefont {C.~R.}\ \bibnamefont {Dean}},
  \bibinfo {author} {\bibfnamefont {J.}~\bibnamefont {Hone}},\ and\ \bibinfo
  {author} {\bibfnamefont {P.~J.}\ \bibnamefont {Schuck}},\ }\href
  {https://doi.org/10.1126/sciadv.abe8691} {\bibfield  {journal} {\bibinfo
  {journal} {Science Advances}\ }\textbf {\bibinfo {volume} {7}},\ \bibinfo
  {pages} {eabe8691} (\bibinfo {year} {2021})}\BibitemShut {NoStop}%
\bibitem [{\citenamefont {Kim}\ \emph {et~al.}(2024)\citenamefont {Kim},
  \citenamefont {Jin}, \citenamefont {Wang}, \citenamefont {He}, \citenamefont
  {Christensen}, \citenamefont {Mele},\ and\ \citenamefont {Zhen}}]{Kim2023b}%
  \BibitemOpen
  \bibfield  {author} {\bibinfo {author} {\bibfnamefont {B.}~\bibnamefont
  {Kim}}, \bibinfo {author} {\bibfnamefont {J.}~\bibnamefont {Jin}}, \bibinfo
  {author} {\bibfnamefont {Z.}~\bibnamefont {Wang}}, \bibinfo {author}
  {\bibfnamefont {L.}~\bibnamefont {He}}, \bibinfo {author} {\bibfnamefont
  {T.}~\bibnamefont {Christensen}}, \bibinfo {author} {\bibfnamefont {E.~J.}\
  \bibnamefont {Mele}},\ and\ \bibinfo {author} {\bibfnamefont
  {B.}~\bibnamefont {Zhen}},\ }\href
  {https://doi.org/10.1038/s41566-023-01318-6} {\bibfield  {journal} {\bibinfo
  {journal} {Nature Photonics}\ }\textbf {\bibinfo {volume} {18}},\ \bibinfo
  {pages} {91} (\bibinfo {year} {2024})}\BibitemShut {NoStop}%
\bibitem [{\citenamefont {Voronin}\ \emph {et~al.}(2024)\citenamefont
  {Voronin}, \citenamefont {Toksumakov}, \citenamefont {Ermolaev},
  \citenamefont {Slavich}, \citenamefont {Tatmyshevskiy}, \citenamefont
  {Novikov}, \citenamefont {Vyshnevyy}, \citenamefont {Arsenin}, \citenamefont
  {Novoselov}, \citenamefont {Ghazaryan}, \citenamefont {Volkov},\ and\
  \citenamefont {Baranov}}]{Voronin2024}%
  \BibitemOpen
  \bibfield  {author} {\bibinfo {author} {\bibfnamefont {K.~V.}\ \bibnamefont
  {Voronin}}, \bibinfo {author} {\bibfnamefont {A.~N.}\ \bibnamefont
  {Toksumakov}}, \bibinfo {author} {\bibfnamefont {G.~A.}\ \bibnamefont
  {Ermolaev}}, \bibinfo {author} {\bibfnamefont {A.~S.}\ \bibnamefont
  {Slavich}}, \bibinfo {author} {\bibfnamefont {M.~K.}\ \bibnamefont
  {Tatmyshevskiy}}, \bibinfo {author} {\bibfnamefont {S.~M.}\ \bibnamefont
  {Novikov}}, \bibinfo {author} {\bibfnamefont {A.~A.}\ \bibnamefont
  {Vyshnevyy}}, \bibinfo {author} {\bibfnamefont {A.~V.}\ \bibnamefont
  {Arsenin}}, \bibinfo {author} {\bibfnamefont {K.~S.}\ \bibnamefont
  {Novoselov}}, \bibinfo {author} {\bibfnamefont {D.~A.}\ \bibnamefont
  {Ghazaryan}}, \bibinfo {author} {\bibfnamefont {V.~S.}\ \bibnamefont
  {Volkov}},\ and\ \bibinfo {author} {\bibfnamefont {D.~G.}\ \bibnamefont
  {Baranov}},\ }\href {https://doi.org/10.1002/lpor.202301113} {\bibfield
  {journal} {\bibinfo  {journal} {Laser \& Photonics Reviews}\ }\textbf
  {\bibinfo {volume} {18}},\ \bibinfo {pages} {2301113} (\bibinfo {year}
  {2024})}\BibitemShut {NoStop}%
\bibitem [{\citenamefont {Verre}\ \emph {et~al.}(2019)\citenamefont {Verre},
  \citenamefont {Baranov}, \citenamefont {Munkhbat}, \citenamefont {Cuadra},
  \citenamefont {Käll},\ and\ \citenamefont {Shegai}}]{Verre2018}%
  \BibitemOpen
  \bibfield  {author} {\bibinfo {author} {\bibfnamefont {R.}~\bibnamefont
  {Verre}}, \bibinfo {author} {\bibfnamefont {D.~G.}\ \bibnamefont {Baranov}},
  \bibinfo {author} {\bibfnamefont {B.}~\bibnamefont {Munkhbat}}, \bibinfo
  {author} {\bibfnamefont {J.}~\bibnamefont {Cuadra}}, \bibinfo {author}
  {\bibfnamefont {M.}~\bibnamefont {Käll}},\ and\ \bibinfo {author}
  {\bibfnamefont {T.}~\bibnamefont {Shegai}},\ }\href
  {https://doi.org/10.1038/s41565-019-0442-x} {\bibfield  {journal} {\bibinfo
  {journal} {Nature Nanotechnology}\ }\textbf {\bibinfo {volume} {14}},\
  \bibinfo {pages} {679} (\bibinfo {year} {2019})}\BibitemShut {NoStop}%
\bibitem [{\citenamefont {Zograf}\ \emph {et~al.}(2024)\citenamefont {Zograf},
  \citenamefont {Polyakov}, \citenamefont {Bancerek}, \citenamefont
  {Antosiewicz}, \citenamefont {Küçüköz},\ and\ \citenamefont
  {Shegai}}]{Zograf2024}%
  \BibitemOpen
  \bibfield  {author} {\bibinfo {author} {\bibfnamefont {G.}~\bibnamefont
  {Zograf}}, \bibinfo {author} {\bibfnamefont {A.~Y.}\ \bibnamefont
  {Polyakov}}, \bibinfo {author} {\bibfnamefont {M.}~\bibnamefont {Bancerek}},
  \bibinfo {author} {\bibfnamefont {T.~J.}\ \bibnamefont {Antosiewicz}},
  \bibinfo {author} {\bibfnamefont {B.}~\bibnamefont {Küçüköz}},\ and\
  \bibinfo {author} {\bibfnamefont {T.~O.}\ \bibnamefont {Shegai}},\ }\href
  {https://doi.org/10.1038/s41566-024-01444-9} {\bibfield  {journal} {\bibinfo
  {journal} {Nature Photonics}\ }\textbf {\bibinfo {volume} {18}},\ \bibinfo
  {pages} {751} (\bibinfo {year} {2024})}\BibitemShut {NoStop}%
\bibitem [{\citenamefont {Munkhbat}\ \emph {et~al.}(2020)\citenamefont
  {Munkhbat}, \citenamefont {Yankovich}, \citenamefont {Baranov}, \citenamefont
  {Verre}, \citenamefont {Olsson},\ and\ \citenamefont {Shegai}}]{Baranov2020}%
  \BibitemOpen
  \bibfield  {author} {\bibinfo {author} {\bibfnamefont {B.}~\bibnamefont
  {Munkhbat}}, \bibinfo {author} {\bibfnamefont {A.~B.}\ \bibnamefont
  {Yankovich}}, \bibinfo {author} {\bibfnamefont {D.~G.}\ \bibnamefont
  {Baranov}}, \bibinfo {author} {\bibfnamefont {R.}~\bibnamefont {Verre}},
  \bibinfo {author} {\bibfnamefont {E.}~\bibnamefont {Olsson}},\ and\ \bibinfo
  {author} {\bibfnamefont {T.~O.}\ \bibnamefont {Shegai}},\ }\href
  {https://doi.org/10.1038/s41467-020-18428-2} {\bibfield  {journal} {\bibinfo
  {journal} {Nature Communications}\ }\textbf {\bibinfo {volume} {11}},\
  \bibinfo {pages} {4604} (\bibinfo {year} {2020})}\BibitemShut {NoStop}%
\bibitem [{\citenamefont {Nauman}\ \emph {et~al.}(2021)\citenamefont {Nauman},
  \citenamefont {Yan}, \citenamefont {Ceglia}, \citenamefont {Rahmani},
  \citenamefont {Kamali}, \citenamefont {Angelis}, \citenamefont
  {Miroshnichenko}, \citenamefont {Lu},\ and\ \citenamefont
  {Neshev}}]{Nauman2021}%
  \BibitemOpen
  \bibfield  {author} {\bibinfo {author} {\bibfnamefont {M.}~\bibnamefont
  {Nauman}}, \bibinfo {author} {\bibfnamefont {J.}~\bibnamefont {Yan}},
  \bibinfo {author} {\bibfnamefont {D.~d.}\ \bibnamefont {Ceglia}}, \bibinfo
  {author} {\bibfnamefont {M.}~\bibnamefont {Rahmani}}, \bibinfo {author}
  {\bibfnamefont {K.~Z.}\ \bibnamefont {Kamali}}, \bibinfo {author}
  {\bibfnamefont {C.~D.}\ \bibnamefont {Angelis}}, \bibinfo {author}
  {\bibfnamefont {A.~E.}\ \bibnamefont {Miroshnichenko}}, \bibinfo {author}
  {\bibfnamefont {Y.}~\bibnamefont {Lu}},\ and\ \bibinfo {author}
  {\bibfnamefont {D.~N.}\ \bibnamefont {Neshev}},\ }\href
  {https://doi.org/10.1038/s41467-021-25717-x} {\bibfield  {journal} {\bibinfo
  {journal} {Nature Communications}\ }\textbf {\bibinfo {volume} {12}},\
  \bibinfo {pages} {5597} (\bibinfo {year} {2021})}\BibitemShut {NoStop}%
\bibitem [{\citenamefont {Weber}\ \emph {et~al.}(2023)\citenamefont {Weber},
  \citenamefont {Kühner}, \citenamefont {Sortino}, \citenamefont {Mhenni},
  \citenamefont {Wilson}, \citenamefont {Kühne}, \citenamefont {Finley},
  \citenamefont {Maier},\ and\ \citenamefont {Tittl}}]{Weber2022a}%
  \BibitemOpen
  \bibfield  {author} {\bibinfo {author} {\bibfnamefont {T.}~\bibnamefont
  {Weber}}, \bibinfo {author} {\bibfnamefont {L.}~\bibnamefont {Kühner}},
  \bibinfo {author} {\bibfnamefont {L.}~\bibnamefont {Sortino}}, \bibinfo
  {author} {\bibfnamefont {A.~B.}\ \bibnamefont {Mhenni}}, \bibinfo {author}
  {\bibfnamefont {N.~P.}\ \bibnamefont {Wilson}}, \bibinfo {author}
  {\bibfnamefont {J.}~\bibnamefont {Kühne}}, \bibinfo {author} {\bibfnamefont
  {J.~J.}\ \bibnamefont {Finley}}, \bibinfo {author} {\bibfnamefont {S.~A.}\
  \bibnamefont {Maier}},\ and\ \bibinfo {author} {\bibfnamefont
  {A.}~\bibnamefont {Tittl}},\ }\href
  {https://doi.org/10.1038/s41563-023-01580-7} {\bibfield  {journal} {\bibinfo
  {journal} {Nature Materials}\ }\textbf {\bibinfo {volume} {22}},\ \bibinfo
  {pages} {970} (\bibinfo {year} {2023})}\BibitemShut {NoStop}%
\bibitem [{\citenamefont {Shen}\ \emph {et~al.}(2022)\citenamefont {Shen},
  \citenamefont {Zhang}, \citenamefont {Zhou}, \citenamefont {Ma},
  \citenamefont {Chen}, \citenamefont {Chen}, \citenamefont {Wang},
  \citenamefont {Xu},\ and\ \citenamefont {Chen}}]{Shen2022b}%
  \BibitemOpen
  \bibfield  {author} {\bibinfo {author} {\bibfnamefont {F.}~\bibnamefont
  {Shen}}, \bibinfo {author} {\bibfnamefont {Z.}~\bibnamefont {Zhang}},
  \bibinfo {author} {\bibfnamefont {Y.}~\bibnamefont {Zhou}}, \bibinfo {author}
  {\bibfnamefont {J.}~\bibnamefont {Ma}}, \bibinfo {author} {\bibfnamefont
  {K.}~\bibnamefont {Chen}}, \bibinfo {author} {\bibfnamefont {H.}~\bibnamefont
  {Chen}}, \bibinfo {author} {\bibfnamefont {S.}~\bibnamefont {Wang}}, \bibinfo
  {author} {\bibfnamefont {J.}~\bibnamefont {Xu}},\ and\ \bibinfo {author}
  {\bibfnamefont {Z.}~\bibnamefont {Chen}},\ }\href
  {https://doi.org/10.1038/s41467-022-33088-0} {\bibfield  {journal} {\bibinfo
  {journal} {Nature Communications}\ }\textbf {\bibinfo {volume} {13}},\
  \bibinfo {pages} {5597} (\bibinfo {year} {2022})}\BibitemShut {NoStop}%
\bibitem [{\citenamefont {Ling}\ \emph {et~al.}(2023)\citenamefont {Ling},
  \citenamefont {Manna}, \citenamefont {Shen}, \citenamefont {Tung},
  \citenamefont {Sharp}, \citenamefont {Fröch}, \citenamefont {Dai},
  \citenamefont {Majumdar},\ and\ \citenamefont {Davoyan}}]{Ling2023}%
  \BibitemOpen
  \bibfield  {author} {\bibinfo {author} {\bibfnamefont {H.}~\bibnamefont
  {Ling}}, \bibinfo {author} {\bibfnamefont {A.}~\bibnamefont {Manna}},
  \bibinfo {author} {\bibfnamefont {J.}~\bibnamefont {Shen}}, \bibinfo {author}
  {\bibfnamefont {H.-T.}\ \bibnamefont {Tung}}, \bibinfo {author}
  {\bibfnamefont {D.}~\bibnamefont {Sharp}}, \bibinfo {author} {\bibfnamefont
  {J.}~\bibnamefont {Fröch}}, \bibinfo {author} {\bibfnamefont
  {S.}~\bibnamefont {Dai}}, \bibinfo {author} {\bibfnamefont {A.}~\bibnamefont
  {Majumdar}},\ and\ \bibinfo {author} {\bibfnamefont {A.~R.}\ \bibnamefont
  {Davoyan}},\ }\href {https://doi.org/10.1364/optica.499059} {\bibfield
  {journal} {\bibinfo  {journal} {Optica}\ }\textbf {\bibinfo {volume} {10}},\
  \bibinfo {pages} {1345} (\bibinfo {year} {2023})}\BibitemShut {NoStop}%
\bibitem [{\citenamefont {Xu}\ \emph {et~al.}(2022)\citenamefont {Xu},
  \citenamefont {Trovatello}, \citenamefont {Mooshammer}, \citenamefont {Shao},
  \citenamefont {Zhang}, \citenamefont {Yao}, \citenamefont {Basov},
  \citenamefont {Cerullo},\ and\ \citenamefont {Schuck}}]{Xu2022a}%
  \BibitemOpen
  \bibfield  {author} {\bibinfo {author} {\bibfnamefont {X.}~\bibnamefont
  {Xu}}, \bibinfo {author} {\bibfnamefont {C.}~\bibnamefont {Trovatello}},
  \bibinfo {author} {\bibfnamefont {F.}~\bibnamefont {Mooshammer}}, \bibinfo
  {author} {\bibfnamefont {Y.}~\bibnamefont {Shao}}, \bibinfo {author}
  {\bibfnamefont {S.}~\bibnamefont {Zhang}}, \bibinfo {author} {\bibfnamefont
  {K.}~\bibnamefont {Yao}}, \bibinfo {author} {\bibfnamefont {D.~N.}\
  \bibnamefont {Basov}}, \bibinfo {author} {\bibfnamefont {G.}~\bibnamefont
  {Cerullo}},\ and\ \bibinfo {author} {\bibfnamefont {P.~J.}\ \bibnamefont
  {Schuck}},\ }\href {https://doi.org/10.1038/s41566-022-01053-4} {\bibfield
  {journal} {\bibinfo  {journal} {Nature Photonics}\ }\textbf {\bibinfo
  {volume} {16}},\ \bibinfo {pages} {698} (\bibinfo {year} {2022})}\BibitemShut
  {NoStop}%
\bibitem [{\citenamefont {Lee}\ \emph {et~al.}(2024)\citenamefont {Lee},
  \citenamefont {Lee}, \citenamefont {Choi}, \citenamefont {Choi},\ and\
  \citenamefont {Gong}}]{Lee2024}%
  \BibitemOpen
  \bibfield  {author} {\bibinfo {author} {\bibfnamefont {S.~W.}\ \bibnamefont
  {Lee}}, \bibinfo {author} {\bibfnamefont {J.~S.}\ \bibnamefont {Lee}},
  \bibinfo {author} {\bibfnamefont {W.~H.}\ \bibnamefont {Choi}}, \bibinfo
  {author} {\bibfnamefont {D.}~\bibnamefont {Choi}},\ and\ \bibinfo {author}
  {\bibfnamefont {S.-H.}\ \bibnamefont {Gong}},\ }\href
  {https://doi.org/10.1038/s41467-024-46701-1} {\bibfield  {journal} {\bibinfo
  {journal} {Nature Communications}\ }\textbf {\bibinfo {volume} {15}},\
  \bibinfo {pages} {2331} (\bibinfo {year} {2024})}\BibitemShut {NoStop}%
\bibitem [{\citenamefont {Sortino}\ \emph {et~al.}(2024)\citenamefont
  {Sortino}, \citenamefont {Biechteler}, \citenamefont {Lafeta}, \citenamefont
  {Kühner}, \citenamefont {Hartschuh}, \citenamefont {Menezes}, \citenamefont
  {Maier},\ and\ \citenamefont {Tittl}}]{Sortino2024}%
  \BibitemOpen
  \bibfield  {author} {\bibinfo {author} {\bibfnamefont {L.}~\bibnamefont
  {Sortino}}, \bibinfo {author} {\bibfnamefont {J.}~\bibnamefont {Biechteler}},
  \bibinfo {author} {\bibfnamefont {L.}~\bibnamefont {Lafeta}}, \bibinfo
  {author} {\bibfnamefont {L.}~\bibnamefont {Kühner}}, \bibinfo {author}
  {\bibfnamefont {A.}~\bibnamefont {Hartschuh}}, \bibinfo {author}
  {\bibfnamefont {L.~d.~S.}\ \bibnamefont {Menezes}}, \bibinfo {author}
  {\bibfnamefont {S.~A.}\ \bibnamefont {Maier}},\ and\ \bibinfo {author}
  {\bibfnamefont {A.}~\bibnamefont {Tittl}},\ }\href@noop {} {\bibfield
  {journal} {\bibinfo  {journal} {arXiv:2407.16480}\ } (\bibinfo {year}
  {2024})}\BibitemShut {NoStop}%
\bibitem [{\citenamefont {Kruk}\ \emph {et~al.}(2022)\citenamefont {Kruk},
  \citenamefont {Wang}, \citenamefont {Sain}, \citenamefont {Dong},
  \citenamefont {Yang}, \citenamefont {Zentgraf},\ and\ \citenamefont
  {Kivshar}}]{Kruk2022}%
  \BibitemOpen
  \bibfield  {author} {\bibinfo {author} {\bibfnamefont {S.~S.}\ \bibnamefont
  {Kruk}}, \bibinfo {author} {\bibfnamefont {L.}~\bibnamefont {Wang}}, \bibinfo
  {author} {\bibfnamefont {B.}~\bibnamefont {Sain}}, \bibinfo {author}
  {\bibfnamefont {Z.}~\bibnamefont {Dong}}, \bibinfo {author} {\bibfnamefont
  {J.}~\bibnamefont {Yang}}, \bibinfo {author} {\bibfnamefont {T.}~\bibnamefont
  {Zentgraf}},\ and\ \bibinfo {author} {\bibfnamefont {Y.}~\bibnamefont
  {Kivshar}},\ }\href {https://doi.org/10.1038/s41566-022-01018-7} {\bibfield
  {journal} {\bibinfo  {journal} {Nature Photonics}\ }\textbf {\bibinfo
  {volume} {16}},\ \bibinfo {pages} {561} (\bibinfo {year} {2022})}\BibitemShut
  {NoStop}%
\bibitem [{\citenamefont {Basov}\ \emph {et~al.}(2020)\citenamefont {Basov},
  \citenamefont {Asenjo-Garcia}, \citenamefont {Schuck}, \citenamefont {Zhu},\
  and\ \citenamefont {Rubio}}]{Basov2020}%
  \BibitemOpen
  \bibfield  {author} {\bibinfo {author} {\bibfnamefont {D.~N.}\ \bibnamefont
  {Basov}}, \bibinfo {author} {\bibfnamefont {A.}~\bibnamefont
  {Asenjo-Garcia}}, \bibinfo {author} {\bibfnamefont {P.~J.}\ \bibnamefont
  {Schuck}}, \bibinfo {author} {\bibfnamefont {X.}~\bibnamefont {Zhu}},\ and\
  \bibinfo {author} {\bibfnamefont {A.}~\bibnamefont {Rubio}},\ }\href
  {https://doi.org/10.1515/nanoph-2020-0449} {\bibfield  {journal} {\bibinfo
  {journal} {Nanophotonics}\ }\textbf {\bibinfo {volume} {10}},\ \bibinfo
  {pages} {549} (\bibinfo {year} {2020})}\BibitemShut {NoStop}%
\bibitem [{\citenamefont {Shen}(1989)}]{Shen1989}%
  \BibitemOpen
  \bibfield  {author} {\bibinfo {author} {\bibfnamefont {Y.}~\bibnamefont
  {Shen}},\ }\href {https://doi.org/10.1146/annurev.physchem.40.1.327}
  {\bibfield  {journal} {\bibinfo  {journal} {Annual Review of Physical
  Chemistry}\ }\textbf {\bibinfo {volume} {40}},\ \bibinfo {pages} {327}
  (\bibinfo {year} {1989})}\BibitemShut {NoStop}%
\bibitem [{\citenamefont {Xu}\ \emph {et~al.}(2018)\citenamefont {Xu},
  \citenamefont {Rahmani}, \citenamefont {Kamali}, \citenamefont
  {Lamprianidis}, \citenamefont {Ghirardini}, \citenamefont {Sautter},
  \citenamefont {Camacho-Morales}, \citenamefont {Chen}, \citenamefont {Parry},
  \citenamefont {Staude}, \citenamefont {Zhang}, \citenamefont {Neshev},\ and\
  \citenamefont {Miroshnichenko}}]{Xu2018a}%
  \BibitemOpen
  \bibfield  {author} {\bibinfo {author} {\bibfnamefont {L.}~\bibnamefont
  {Xu}}, \bibinfo {author} {\bibfnamefont {M.}~\bibnamefont {Rahmani}},
  \bibinfo {author} {\bibfnamefont {K.~Z.}\ \bibnamefont {Kamali}}, \bibinfo
  {author} {\bibfnamefont {A.}~\bibnamefont {Lamprianidis}}, \bibinfo {author}
  {\bibfnamefont {L.}~\bibnamefont {Ghirardini}}, \bibinfo {author}
  {\bibfnamefont {J.}~\bibnamefont {Sautter}}, \bibinfo {author} {\bibfnamefont
  {R.}~\bibnamefont {Camacho-Morales}}, \bibinfo {author} {\bibfnamefont
  {H.}~\bibnamefont {Chen}}, \bibinfo {author} {\bibfnamefont {M.}~\bibnamefont
  {Parry}}, \bibinfo {author} {\bibfnamefont {I.}~\bibnamefont {Staude}},
  \bibinfo {author} {\bibfnamefont {G.}~\bibnamefont {Zhang}}, \bibinfo
  {author} {\bibfnamefont {D.}~\bibnamefont {Neshev}},\ and\ \bibinfo {author}
  {\bibfnamefont {A.~E.}\ \bibnamefont {Miroshnichenko}},\ }\href
  {https://doi.org/10.1038/s41377-018-0051-8} {\bibfield  {journal} {\bibinfo
  {journal} {Light: Science \& Applications}\ }\textbf {\bibinfo {volume}
  {7}},\ \bibinfo {pages} {44} (\bibinfo {year} {2018})}\BibitemShut {NoStop}%
\bibitem [{\citenamefont {Purdie}\ \emph {et~al.}(2018)\citenamefont {Purdie},
  \citenamefont {Pugno}, \citenamefont {Taniguchi}, \citenamefont {Watanabe},
  \citenamefont {Ferrari},\ and\ \citenamefont {Lombardo}}]{Purdie2018}%
  \BibitemOpen
  \bibfield  {author} {\bibinfo {author} {\bibfnamefont {D.~G.}\ \bibnamefont
  {Purdie}}, \bibinfo {author} {\bibfnamefont {N.~M.}\ \bibnamefont {Pugno}},
  \bibinfo {author} {\bibfnamefont {T.}~\bibnamefont {Taniguchi}}, \bibinfo
  {author} {\bibfnamefont {K.}~\bibnamefont {Watanabe}}, \bibinfo {author}
  {\bibfnamefont {A.~C.}\ \bibnamefont {Ferrari}},\ and\ \bibinfo {author}
  {\bibfnamefont {A.}~\bibnamefont {Lombardo}},\ }\href
  {https://doi.org/10.1038/s41467-018-07558-3} {\bibfield  {journal} {\bibinfo
  {journal} {Nature Communications}\ }\textbf {\bibinfo {volume} {9}},\
  \bibinfo {pages} {5387} (\bibinfo {year} {2018})}\BibitemShut {NoStop}%
\bibitem [{\citenamefont {Franceschini}\ \emph {et~al.}(2023)\citenamefont
  {Franceschini}, \citenamefont {Tognazzi}, \citenamefont {Finco},
  \citenamefont {Carletti}, \citenamefont {Alessandri}, \citenamefont {Cino},
  \citenamefont {Angelis}, \citenamefont {Takayama}, \citenamefont {Malureanu},
  \citenamefont {Lavrinenko},\ and\ \citenamefont {Ceglia}}]{Franceschini2023}%
  \BibitemOpen
  \bibfield  {author} {\bibinfo {author} {\bibfnamefont {P.}~\bibnamefont
  {Franceschini}}, \bibinfo {author} {\bibfnamefont {A.}~\bibnamefont
  {Tognazzi}}, \bibinfo {author} {\bibfnamefont {G.}~\bibnamefont {Finco}},
  \bibinfo {author} {\bibfnamefont {L.}~\bibnamefont {Carletti}}, \bibinfo
  {author} {\bibfnamefont {I.}~\bibnamefont {Alessandri}}, \bibinfo {author}
  {\bibfnamefont {A.~C.}\ \bibnamefont {Cino}}, \bibinfo {author}
  {\bibfnamefont {C.~D.}\ \bibnamefont {Angelis}}, \bibinfo {author}
  {\bibfnamefont {O.}~\bibnamefont {Takayama}}, \bibinfo {author}
  {\bibfnamefont {R.}~\bibnamefont {Malureanu}}, \bibinfo {author}
  {\bibfnamefont {A.~V.}\ \bibnamefont {Lavrinenko}},\ and\ \bibinfo {author}
  {\bibfnamefont {D.~d.}\ \bibnamefont {Ceglia}},\ }\href
  {https://doi.org/10.1063/5.0159275} {\bibfield  {journal} {\bibinfo
  {journal} {Applied Physics Letters}\ }\textbf {\bibinfo {volume} {123}},\
  \bibinfo {pages} {071701} (\bibinfo {year} {2023})}\BibitemShut {NoStop}%
\end{thebibliography}%
\bibliographystyle{apsrev4-2}

\bigskip
\noindent
\textbf{Acknowledgements}

\noindent
This work was funded by the Deutsche Forschungsgemeinschaft (DFG, German Research Foundation) under Germany’s Excellence Strategy (EXC 2089/1 – 390776260), Sachbeihilfe MA 4699/7-1 and the Emmy Noether program (TI 1063/1); the Bavarian program Solar Energies Go Hybrid (SolTech) and the Center for NanoScience (CeNS). L.S. acknowledges funding support through a Humboldt Research Fellowship from the Alexander von Humboldt Foundation. A.To. acknowledges the financial support from the European Union through "FESR o FSE, PON Ricerca e Innovazione 2014-2020 - DM 1062/2021" and the University of Palermo through "Fondo Finalizzato alla Ricerca di Ateneo 2024 (FFR2024)". This work was partially supported by the European Union under the Italian National Recovery and Resilience Plan (NRRP) of NextGenerationEU, of partnership on “Telecommunications of the Future” (PE00000001 - program “RESTART”), S2 SUPER – Programmable Networks, Cascade project PRISM - CUP: E13C22001870001.
C.D.A. and P.F. acknowledge the financial support from the European Union "METAFAST" H2020-FETOPEN-2018-2020 project, grant agreement no. 899673; from Ministero Italiano dell'Istruzione (MIUR) through the "METEOR" project PRIN-2020 2020EY2LJT\_002. A.Ti. further acknowledges funding support through the European Union (ERC, METANEXT, 101078018). Views and opinions expressed are however those of the author(s) only and do not necessarily reflect those of the European Union or the European Research Council Executive Agency. Neither the European Union nor the granting authority can be held responsible for them. 

\bigskip
\noindent
\textbf{Author Contributions}

\noindent
A.To., L.S, P.F., C.D.A. conceived the idea. J.B. fabricated the samples. E.B. perfomed the AFM analysis. A.To. and P.F. carried out optical spectroscopy experiments. L.S., A.To, P.F. analyzed the data with contributions from all authors. L.S. wrote the manuscript with contributions from all authors. L.S, C.D.A, A.C.C. and A.Ti. supervised various aspects of the project. 

\bigskip
\noindent
\textbf{Conflict of interest}

\noindent
The authors declare no competing interests.

\clearpage

\renewcommand{\figurename}{SUPPLEMENTARY FIGURE}
\setcounter{figure}{0}   
\linespread{1}

\section*{Supplementary information for: Interface second harmonic generation enhancement in hetero-bilayer van der Waals nanoantennas }
\begin{center}
A. Tognazzi \textit{et al.}
\end{center}

\newpage

\section*{Supplementary note I: Optical images of the double-layer TMDC nanoantennas}

\begin{figure}[h!]
	\centering
	\includegraphics[width=0.8\linewidth]{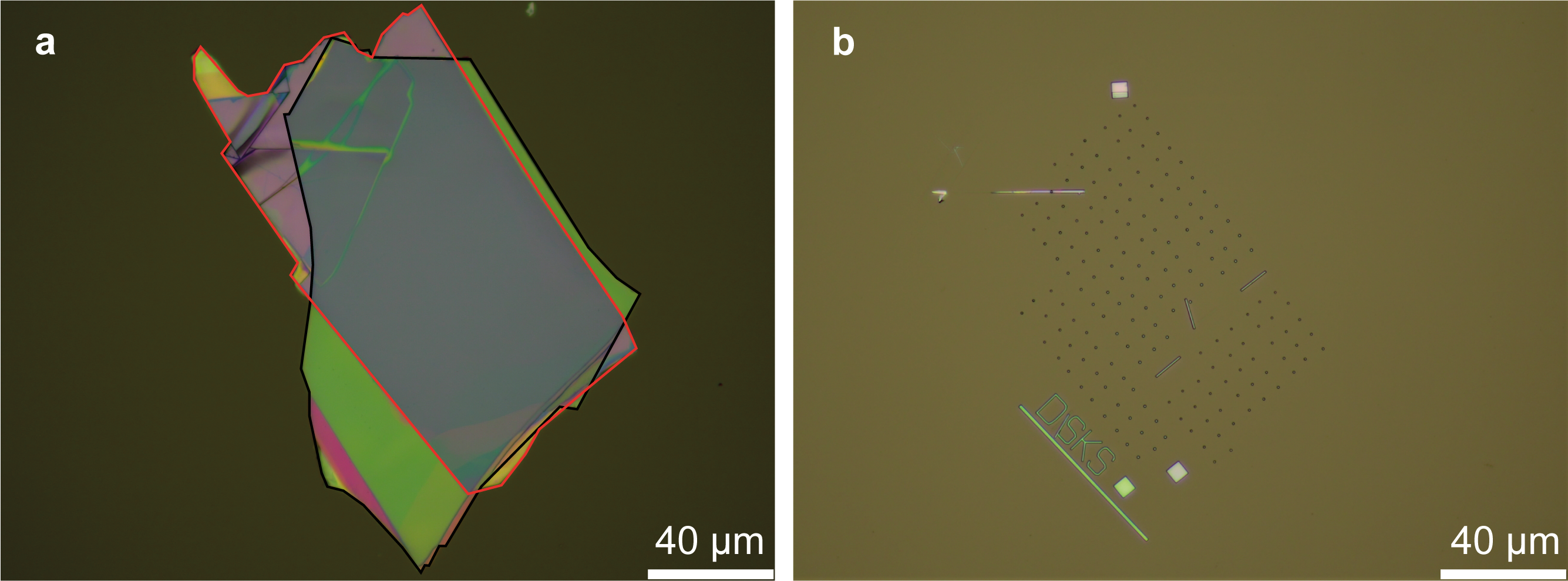}
	\caption{(a) Bright field microscopy image of the fabricated stack before nanofabrication. (b) Image of the final sample after nanofabrication.}
	\label{fig:si1}
\end{figure}

\section*{Supplementary note II: AFM height profiles}

\begin{figure}[h!]
	\centering
	\includegraphics[width=0.5\linewidth]{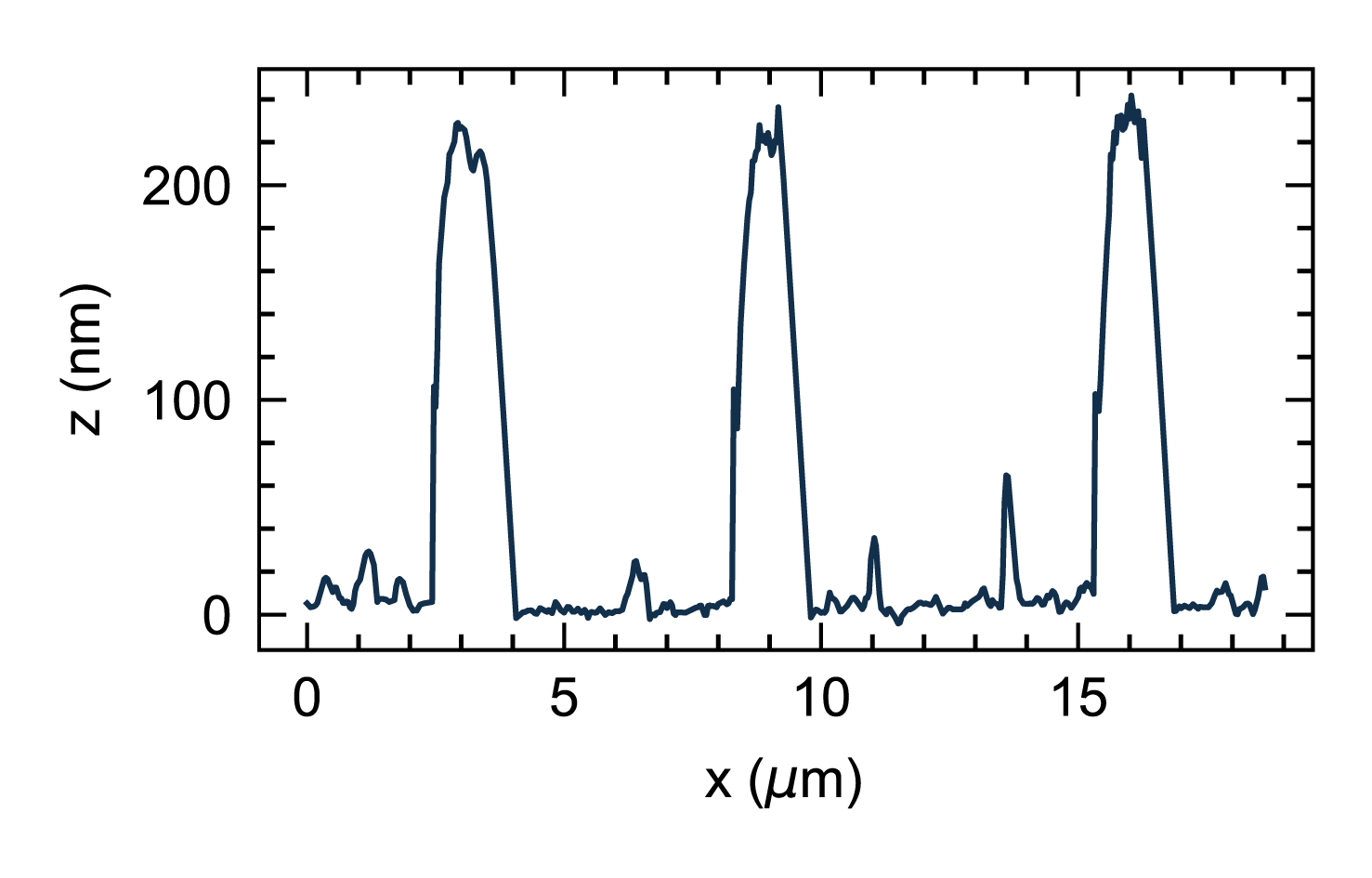}
	\caption{Atomic force microscopy profiles of three fabricated nanoantennas. The thickness of the final nanostructures is consistent with the pre-fabrication observations, confirming that the overetching has not affected the height of the TMDC hetero-bilayer.}
	\label{fig:si_afm}
\end{figure}

\clearpage
\section*{Supplementary note III: Linear simulations of hexagonal nanostructures}

\noindent
We verify that the spectral position of the anapole is not dependent upon the impinging polarization angle $\theta$ by performing linear simulations with Comsol Multiphysics  as shown in Supplementary Figure~\ref{fig:si_linear_pol_dep}, with the dielectric constant shown in Figure~2 in the main text. We illuminate the nanoantenna with a normally incident monochromatic wave and ensure to remove any reflection from the domain boundaries by introducing perfectly matched layers and scattering boundary conditions. Although the electric field distribution is perturbed when the polarization angle is changed (see Supplementary Figure~\ref{fig:si_pol_field}) the spectral position of the anapole condition is not significantly affected.

\begin{figure}[h!]
	\centering
	\includegraphics[width=0.8\linewidth]{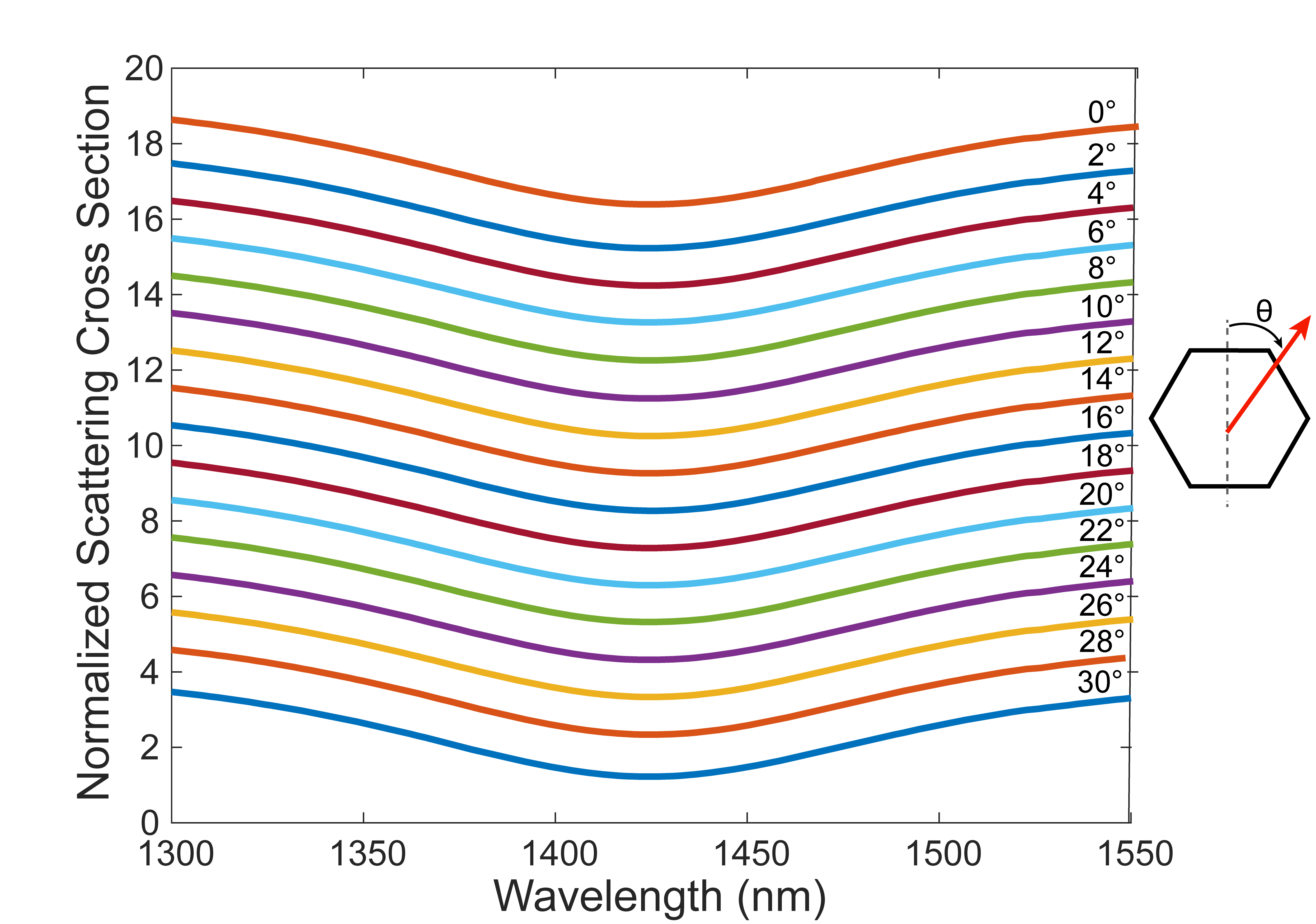}
	\caption{Simulated normalized scattering cross section at different polarization angles ($\theta$) of a disk with radius of 300~nm. The red arrow in the right sketch represents the polarization direction. The traces are vertically shifted for sake of visualization.}
	\label{fig:si_linear_pol_dep}
\end{figure}

\begin{figure}[h!]
	\centering
	\includegraphics[width=0.9\linewidth]{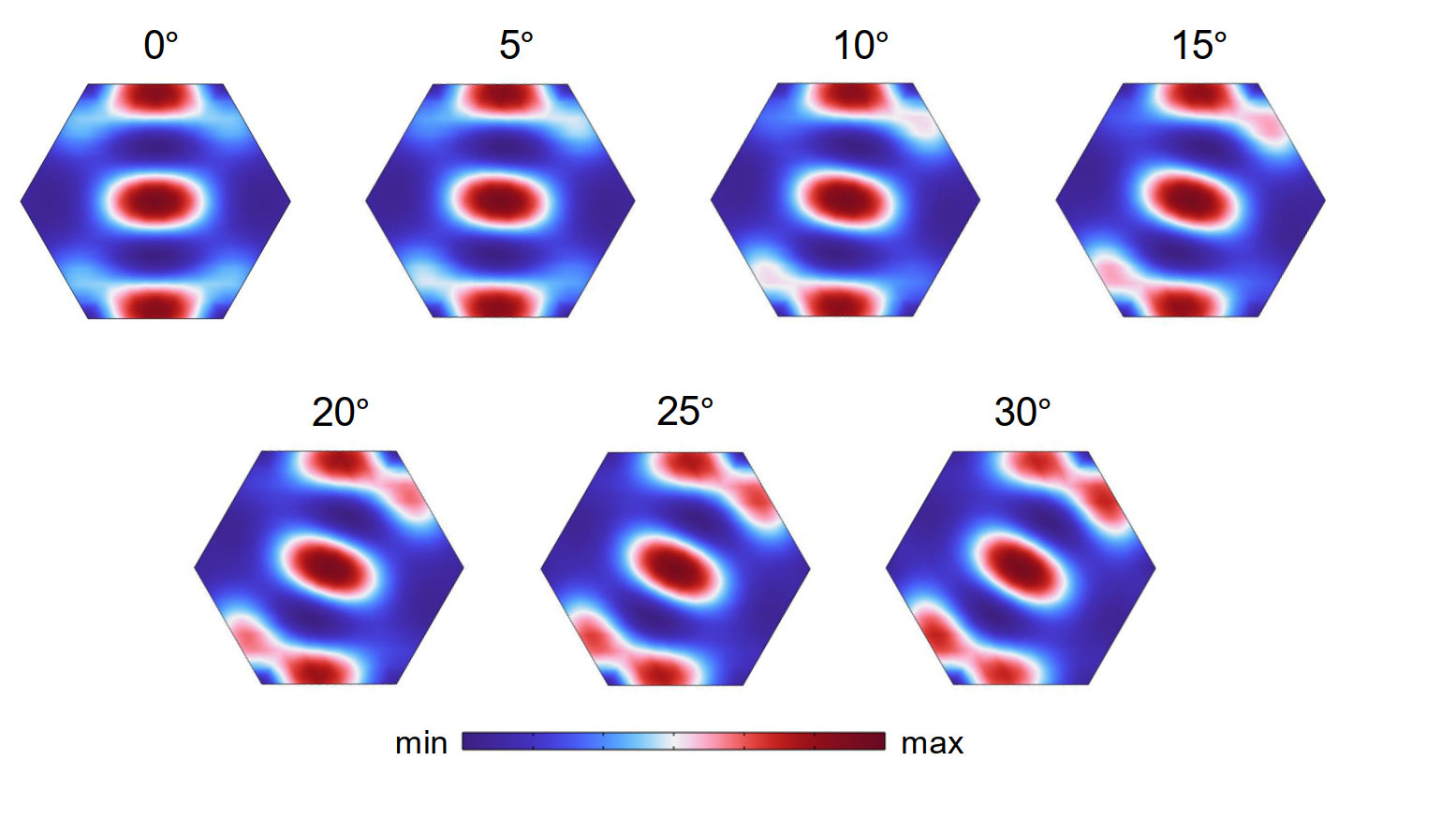}
	\caption{Simulated electromagnetic field enhancement ($|E|/|E_0|)^2$ at different polarization angles at 1420~nm for a disk with radius 300~nm.}
	\label{fig:si_pol_field}
\end{figure}

\begin{figure}[h!]
	\centering
	\includegraphics[width=0.9\linewidth]{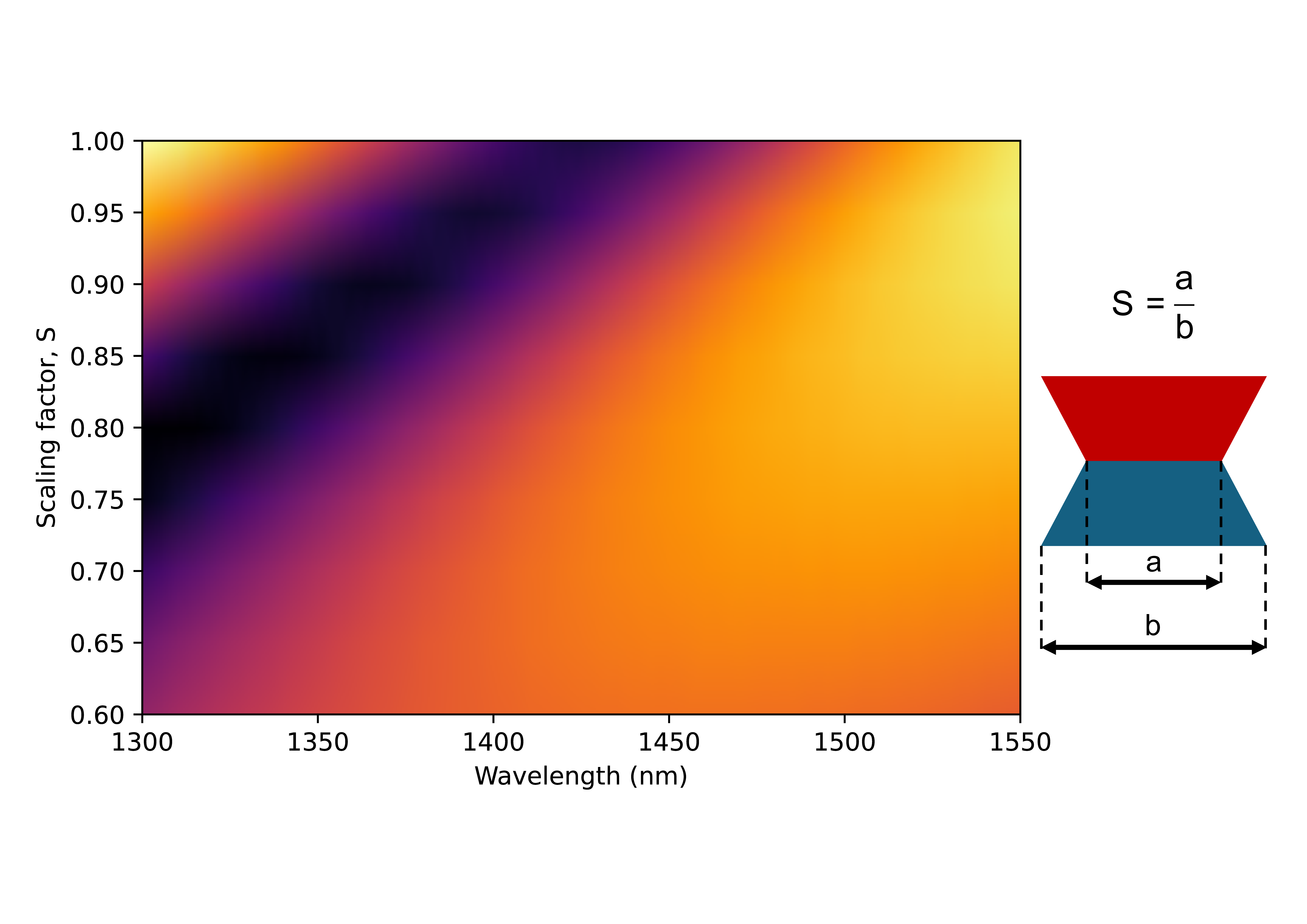}
	\caption{Simulated normalized scattering cross section of tilted sidewalls nanoantennas, as a function of the wavelength and the scaling factor  ($b=300$~nm).}
	\label{fig:si_scaling_factor}
\end{figure}

\clearpage
\section*{Supplementary note IV: Linear visible reflectance of WS$_2$/MoS$_2$ nanoantennas}

\begin{figure}[h!]
    \centering
    \includegraphics[width=0.5\linewidth]{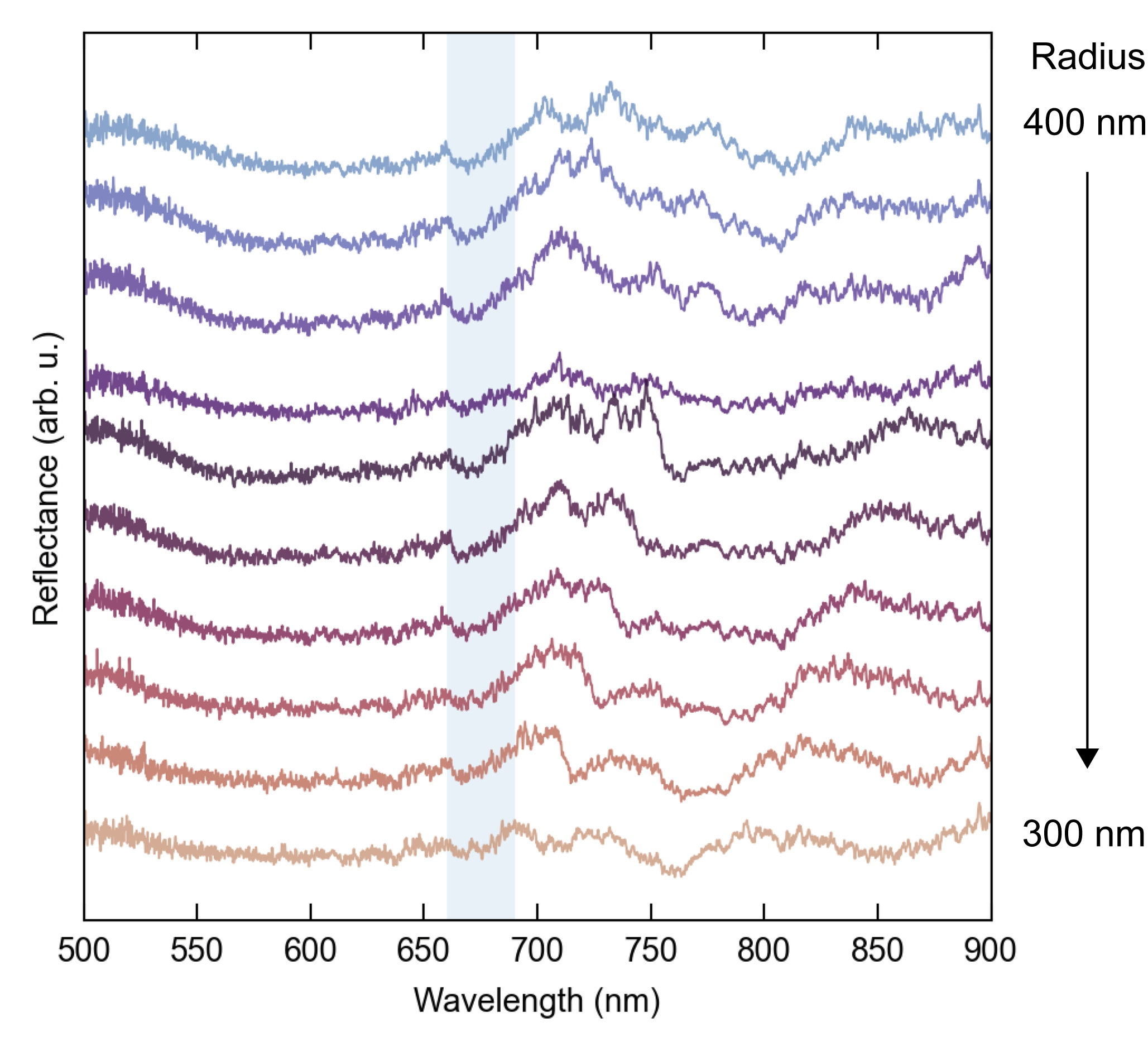}
    \caption{Visible reflectance of WS$_2$/MoS$_2$ nanoantennas with radius ranging from 400~nm to 300~nm. The shaded blue area indicates the position of the $X^{A}_{MoS_2}$ exciton resonance. }
    \label{fig:SI_visantenna}
\end{figure}

\clearpage
\section*{Supplementary note V: SHG normalization}

The experimental approach to SHG measurements was to compare the nonlinear signal intensity with an unpatterned patch, much larger than the laser radius of approximately 1~$\mu$m in diameter. To compare with the SHG counts from the individual nanoantenna, with size smaller than the laser spotsize, we normalized the effective power on the sample by normalizing the values of the SHG counts, obtaining the SHG ratio, defined as: 

\begin{equation}
	SHG_{ratio} = \dfrac{I_{ref}}{A_{laser}}\dfrac{A_r}{I_r}
\end{equation}
\noindent
where $I_{ref}$ is the SHG intensity collected from the reference sample, $A_{laser}$ is the laser area, $I_r$ the SHG intensity collected from the nanoantennas with radius \textit{r}, and $A_r$ is the relative hexagonal area. In Table~\ref{fig:t1} are reported the calculated hexagonal cross section area, and the relative ratio with the laser spot size area of 1 micron in diameter used in the estimation of the SHG ratio in the main text.

\begin{table}[h]
	\begin{tabular}{|l|l|l|}
		\hline
		\multicolumn{1}{|c|}{\textbf{Hexagon   Radius (nm)}} & \multicolumn{1}{c|}{\textbf{Hexagon Area (nm²)}} & \multicolumn{1}{c|}{\textbf{Relative Ratio}} \\ \hline
		260 & 175,902 & 0.224\\ \hline
		270 & 189,122 & 0.241\\ \hline
		280 & 202,857 & 0.258\\ \hline
		290 & 217,109 & 0.276\\ \hline
		300 & 231,878 & 0.295\\ \hline
		310 & 247,163 & 0.315\\ \hline
		320 & 262,964 & 0.335\\ \hline
		330 & 279,281 & 0.356\\ \hline
		340 & 296,115 & 0.377\\ \hline
		350 & 313,465 & 0.399\\ \hline
		360 & 331,331 & 0.422\\ \hline
	\end{tabular}
	\caption{Area of an hexagon for different radial size, defined as the distance from the center to a vertex, and the ratio with the area of a circle of 1 micron in diameter.}
	\label{fig:t1}
\end{table}
\clearpage

\clearpage
\section*{Supplementary note VI: Visible linear reflectance of fabricated nanostructures}

\begin{figure}[h!]
	\centering
	\includegraphics[width=0.6\linewidth]{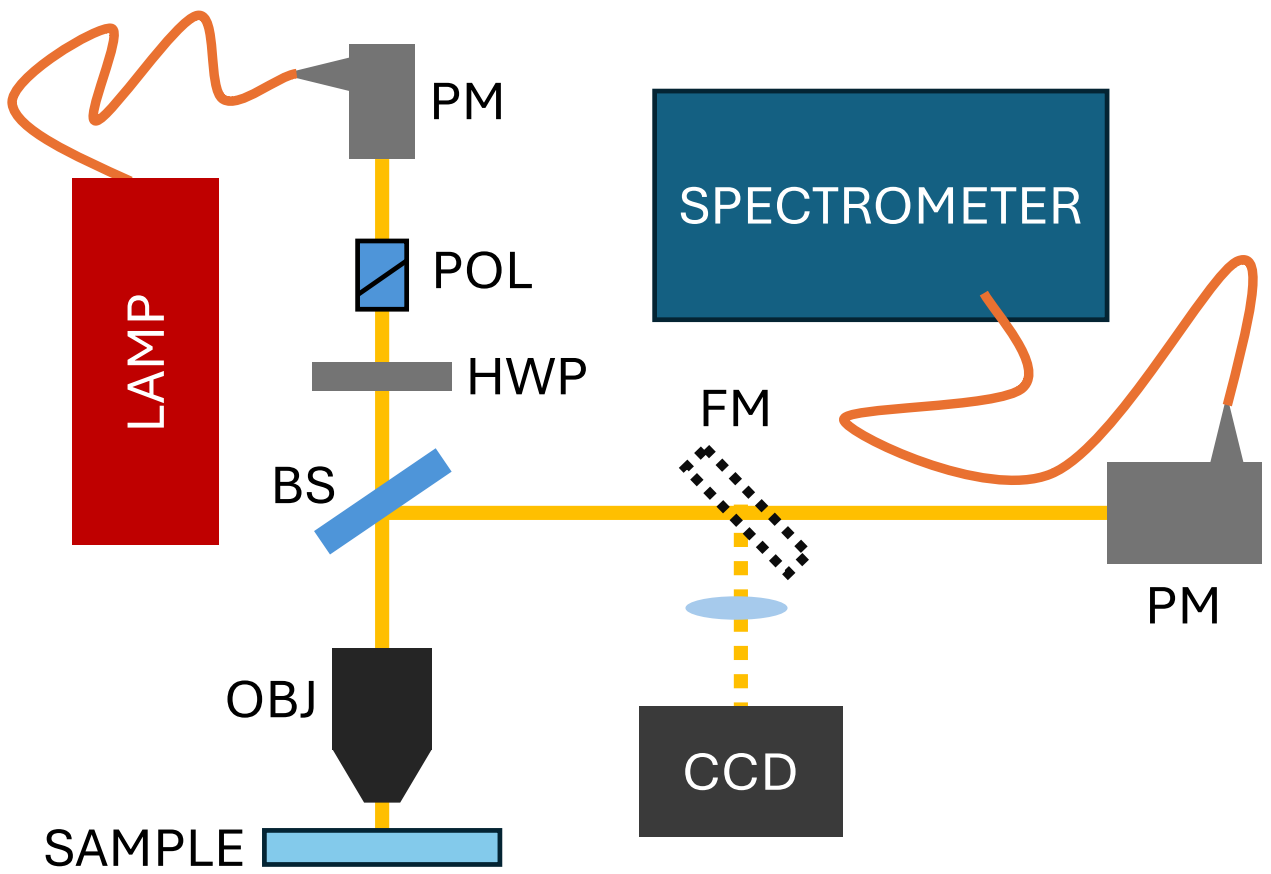}
	\caption{Linear spectroscopy setup. PM: Parabolic mirror, POL: Polarizer, HWP: Half-wave plate, BS: Beam splitter, OBJ: objective, FM: Flip mirror.}
	\label{fig:si_linear}
\end{figure}

\section*{Supplementary note VII: Nonlinear Second Harmonic generation microscopy}

\begin{figure}[h!]
	\centering
	\includegraphics[width=0.8\linewidth]{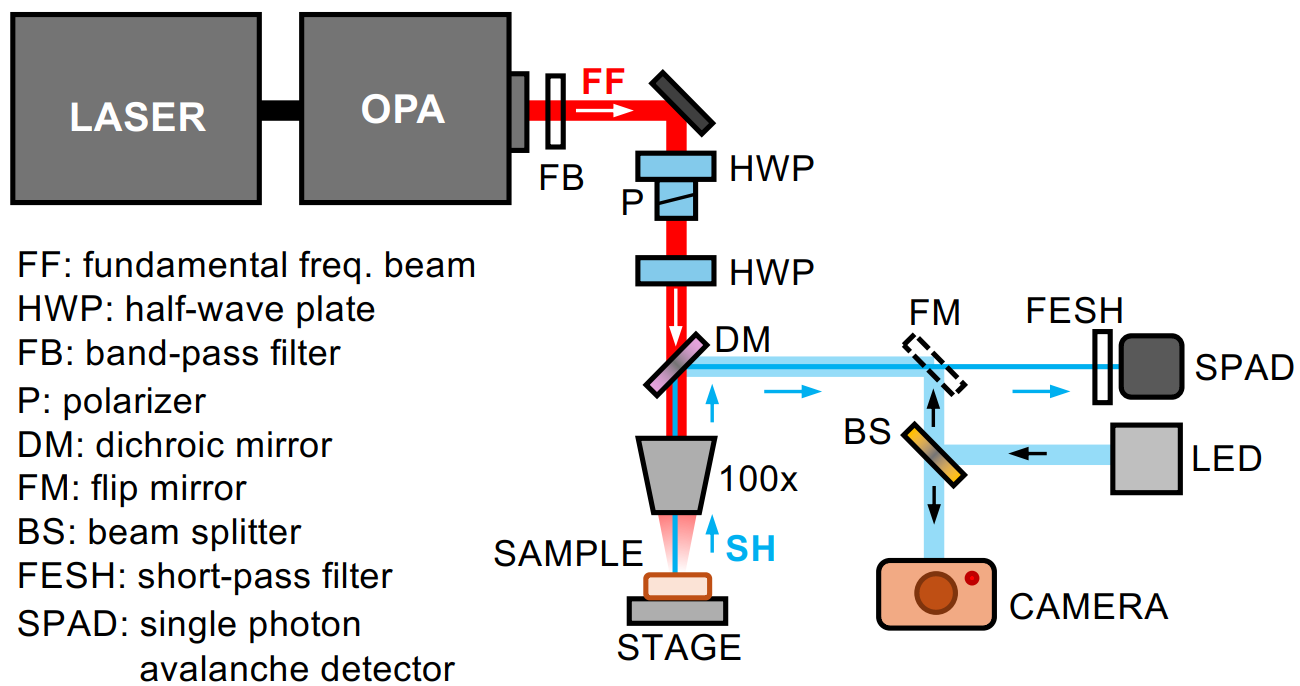}
	\caption{Nonlinear spectroscopy setup.}
	\label{fig:si_nonlinear}
\end{figure}

\end{document}